\documentclass[11pt,twoside]{article}
\usepackage{amsmath,amsthm,amssymb,amsbsy,bm,calc,fancyhdr,fixltx2e,
            float,glossaries,ifthen,mathrsfs,textgreek,titlesec,tocloft,xy}
\input epsf
\PassOptionsToPackage{ctagsplt,Righttag}{amstex}
\usepackage{cite}
\usepackage[dvips,dvipsnames,usenames]{color} 
\usepackage[colorlinks=true,citecolor=MidnightBlue,linkcolor=BrickRed,breaklinks=true]{hyperref} 
\usepackage[usenames,dvipsnames]{xcolor}
\usepackage[super]{nth}
\usepackage{graphicx}
\usepackage{amsmath}
\usepackage{mathtools}
\usepackage{wrapfig}
\usepackage[font={footnotesize}]{caption}
\usepackage{tikz}
\usepackage{hyperref}

\definecolor{darkgreen}{rgb}{0,0.65,0}
\definecolor{darkred}{rgb}{0.65,0,0}
\include{Commands}

\newcommand{\ab}{\mathbf{a}}
\newcommand{\bb}{\mathbf{b}}

\newcommand{\db}{\mathbf{d}}

\newcommand{\fb}{\mathbf{f}}

\newcommand{\mb}{\mathbf{m}}
\newcommand{\nb}{\mathbf{n}}

\newcommand{\vb}{\mathbf{v}}

\newcommand{\xb}{\mathbf{x}}

\newcommand{\Ab}{\mathbf{A}}

\newcommand{\Cb}{\mathbf{C}}

\newcommand{\Gb}{\mathbf{G}}
\newcommand{\Hb}{\mathbf{H}}
\newcommand{\Ib}{\mathbf{I}}
\newcommand{\Jb}{\mathbf{J}}

\newcommand{\Lb}{\mathbf{L}}

\newcommand{\Sb}{\mathbf{S}}
\newcommand{\Tb}{\mathbf{T}}

\newcommand{\Yb}{\mathbf{Y}}

\newcommand{\omegab}{\boldsymbol{\omega}}

\def\f#1{Fig.~\ref{#1}}

\def\k{k_{\rm B}}
\def\kt{k_{\rm B}T}

\advance\textheight by \topskip
\allowdisplaybreaks
\begin{document}


\begin{center}
	{\textbf{\Large{Statistical mechanics of transport processes in active fluids: Equations of hydrodynamics}}
	} \\
	\vspace{0.2in}
	Katherine Klymko$^{1, \ddag}$, Dibyendu Mandal$^{2, \S}$, and Kranthi K. Mandadapu$^{3,4 \dag}$ \\
	\vspace{0.1in}
		\footnotesize{
		{
			$^1$ Department of Chemistry, University of California at Berkeley,
			Berkeley, CA, 94720, USA \\
			$^2$ Department of Physics, University of California at Berkeley,
			Berkeley, CA, 94720, USA \\
			$^3$ Department of Chemical \& Biomolecular Engineering, University of California at Berkeley,\\
			Berkeley, CA, 94720, USA \\
			$^4$ Chemical Sciences Division, Lawrence Berkeley National Laboratory, 
			Berkeley, CA, 94720, USA \\
		}
	}
\end{center}

\begin{abstract}
	The equations of hydrodynamics including mass, linear momentum, angular momentum and energy are derived by coarse-graining the microscopic equations of motion for systems consisting of rotary dumbbells driven by internal torques. 
\end{abstract}

\noindent\rule{4.6cm}{0.4pt}

\noindent \small{\ddag \, katieklymko@gmail.com\\
	\S \, dibyendu.mandal@berkeley.edu \\
	\dag \, kranthi@berkeley.edu
}

\vspace{2pt}

\small \tableofcontents

\section{Introduction} \label{sec:intro}
In this paper, we consider the hydrodynamics of active fluids consisting of structured particles subjected to internal torques (or couples). The system under consideration consists of dumbbells immersed in a fluid and rotated by an equal and opposite force perpendicular to the axis connecting the two ends of the dumbbell. For such active fluids, we derive the associated balances of mass, linear momentum, angular momentum, moment of inertia, total energy, and internal energy by systematically coarse-graining the microscopic equations of motion governing the dynamics of the active particles. We follow the Iriving-Kirkwood procedure  \cite{irving1950statistical}, which was originally used to derive the equations of hydrodynamics of simple fluids. 

In deriving the balance of linear momentum, we find that the symmetry of the stress tensor is broken due to the presence of non-zero torques on individual particles. The broken symmetry of the stress tensor induces internal  spin in the fluid and leads us to consider the balance of internal angular momentum in addition to the usual moment of momentum. In the absence of spin, the moment of momentum is the same as the total angular momentum. In deriving the form of the balance of total angular momentum, we find the microscopic expressions for the couple stress tensor that drives the spin field. We show that the couple stresses contains contributions from both intermolecular interactions and the active forces. The presence of spin leads to the idea of balance of moment of inertia due to the constant exchange of particles in a small neighborhood around a macroscopic point. We derive the associated balance of moment of inertia at the macroscale and identify the moment of inertia flux that induces its transport. Finally, we obtain the balances of total and internal energy of the active fluid and identify the sources of heat and heat fluxes in the system. 

{Our system falls under the general field of active matter \cite{Ramaswamy_2010, Romanczuk_2012, Marchetti_2013,Yeomans_2014,Menzel_2015,Bechinger_2016}, a term used to describe individual particles capable of self-propelled motion. Active matter systems are known to exhibit non-equilibrium phase behaviors with dynamic clustering of active particles \cite{Tailleur_2008,Fily_2012,Buttinoni_2013,Redner_2013,Mognetti_2013,Cates_2015}. Suspensions of active matter also lead to anomalous thermal and mechanical properties with enhanced diffusion \cite{Wu_2000, Leptos_2009}, odd and vanishing viscosities \cite{banerjee2017odd, lopez2015} among many others \cite{Narayan_2007,bialke2015negative,Kaiser_2014,Mallory_2014_PRE_II}.  Of particular recent interest is the concept of pressure in these systems \cite{Takatori_2014_PRL,Solon_2015,Winkler_2015,Speck_2016_PRE,Joyeux_2016,Nikola_2016,Joyeux_2017,Marconi_2017_arXiv,Fily_2017,Sandford_2017_arXiv}, where it is argued that pressure behaves as a state function only for the case of spherical, torque-less active particles but not for general active fluids \cite{Solon_2015}. 

Pressure, being a fundamentally mechanical concept, depends on the nature of the stress tensors arising out of the balance of linear momentum. Mechanically, pressure can be defined as the negative trace of the stress tensor and much microscopic understanding can be gained by knowing the expressions for the stress tensor in terms of the molecular interactions. Microscopic expressions for the stress tensors have been developed for passive systems starting with Clausius \cite{clausius1870}, and finally with the theory proposed by Irving and Kirkwood \cite{irving1950statistical}. The latter work, in particular, obtained the expressions for the stress tensor by deriving the equations of hydrodynamics using the principles of classical statistical mechanics. These expressions explicitly showed the role of intermolecular forces in the stress tensor. Recently, there has been an attempt to extend the theory of Irving and Kirkwood towards active systems consisting of self-propelled Brownian particles \cite{Steffenoni2017}, where the derivations are restricted to the case of balances of mass and linear momentum. However, one may think of active systems consisting of particles with internal structure, the simplest example being dumbbells rotated by force couples or internal torques. These systems introduce an additional concept of spin resulting from the coarse grained angular momentum of the particles. The concept of spin requires development of the balance of angular momentum with the presence of surface couples and the associated couple stress tensor. In these cases, at least for passive particles, it is well known that both angular momentum and linear momentum relations are coupled to each other \cite{Dahler_1961,Dahler_1963,deGroot_1984,Stokes_1966,Stokes_1984,Spencer_2004}. Analogous continuum theories have been used to model active systems made up of particles driven by internal torque \cite{lau2009fluctuating,Stark_2005,Tsai2005,van2016spatiotemporal}. However, rigorous derivations for the expressions of the stress and couple stress tensors in terms of molecular variables are lacking for the case of active systems, and therefore there is an incomplete understanding of the role of active forces. Moreover, equations concerning the balance of energy have not been explored to understand the sources of heat and heat fluxes at the continuum level in such active systems. Our work extends the theory proposed by Irving and Kirkwood to derive the equations of hydrodynamics for systems consisting of active dumbbell particles with expressions for the stress tensor, couple stress tensor, and heat fluxes in terms of molecular variables, thereby providing a molecular basis for understanding the generalized hydrodynamics of active polar suspensions.

Our paper is organized as follows. In Section~\ref{sec:ik_micro}, we describe the microscopic dynamics of the active dumbbell system, and in Sections~\ref{sec:ik_relations}-\ref{sec:ik_energy}, we perform the coarse-graining procedure to derive the balance laws for the active fluid. A concise report of what has been done in this paper is presented in \cite{Mandal2017short}.

\section{Irving-Kirkwood Procedure} \label{sec:irving_kirkwood}

In this section, we derive the balances of mass, linear momentum, angular momentum, moment-of-intertia, energy, and internal energy for systems consisting of active rotating dumbbells. As mentioned before, we follow the procedure of Irving and Kirkwood \cite{irving1950statistical}, where equations of hydrodynamics were derived in the absence of any internal rotation. 

\subsection{Microscopic Equations of Motion}\label{sec:ik_micro}

Our model is a simple active matter system consisting of underdamped dumbbell-like particles where the `atoms' are tethered together harmonically \cite{van2016spatiotemporal} (see \f{fig:dumbbell}). They move according to the following equations of motion:
\begin{equation} \label{eq:motion}
\begin{split}
        \dot{\bold{x}}^{\alpha}_i &= \bold{p}^{\alpha}_i/m_i \ , \\
        \dot{\bold{p}}^{\alpha}_i  & = - \zeta \frac{\bold{p}^{\alpha} _i}{m_i} +  \sum_{j, \beta} \bold{F}^{\alpha \beta}_{ij} +\bold{f}^{\alpha}_i-\frac{\partial u_s(\bold{x}^{1}_i, \bold{x}^{2}_i )}{\partial \bold{x}^{\alpha}_i} +\sqrt{2\kt \zeta}\, \frac{\mathrm{d}\bold{W}}{\mathrm{d}t} \ ,
\end{split}
\end{equation}
\begin{figure}[!t]
	\centering
		\includegraphics[width=0.8\textwidth]{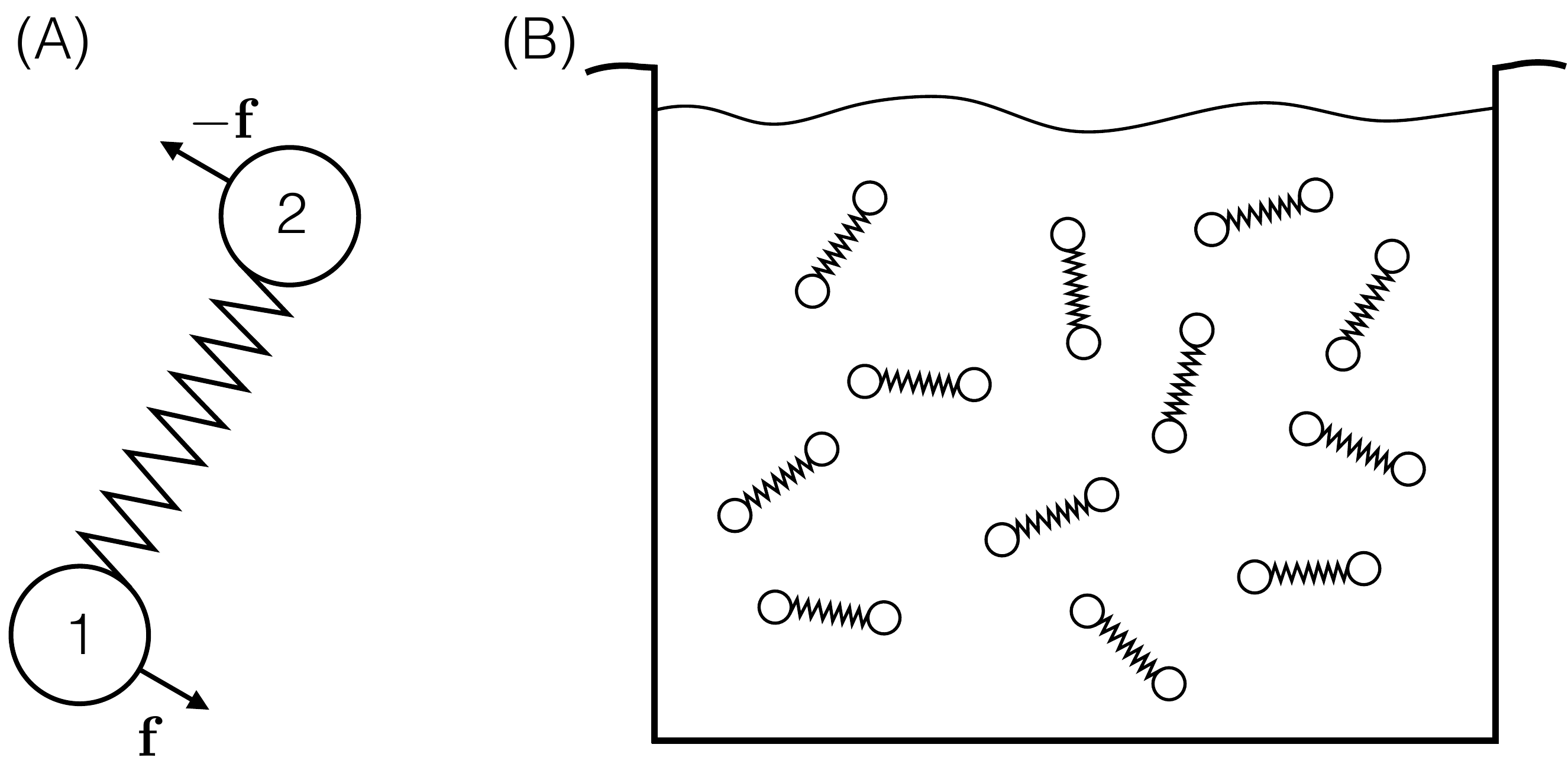}
	\captionsetup{width=0.80\textwidth}
	\caption{
		Active dumbbell particles: (A) A schematic showing an active dumbbell particle with equal and opposite forces on the atoms of the dumbbell. It is assumed that the forces $\fb$ always act perpendicular to the bond connecting the two atoms. (B) A schematic of a fluid consisting of many active dumbbell particles. 
	}
	\label{fig:dumbbell}
\end{figure}
$\\$where $i$ refers to the molecule, $\alpha$ refers to the atom of a molecule with $\alpha \in \{1,2\}$, $\bold{x}^{\alpha}_i$ and $\bold{p}^{\alpha}_i$ are the position and momentum of atom $\alpha$ of molecule $i$, $\zeta$ is the drag coefficient, $\k$ is Boltzmann's constant, $T$ is the temperature of the bath, and $\dfrac{\mathrm{d}\bold{W}}{\mathrm{d}t}$ is Gaussian white noise \cite{zwanzig2001nonequilibrium}.  The term $u_s(\bold{x}^{1}_i, \bold{x}^{2}_i )$ is the harmonic spring energy connecting atoms 1 and 2 of molecule $i$. Also, $\bold{F}^{\alpha \beta}_{ij}$ is the force on atom $\alpha$ of molecule $i$ due to atom $\beta$ of molecule $j$ given by $\bold{F}^{\alpha \beta}_{ij} = -\dfrac{\partial u_2(\bold{x}^{\alpha}_i,\bold{x}^{\beta}_j)}{\partial \bold{x}^{\alpha}_{i}}$, where $u_2(\bold{x}^{\alpha}_i,\bold{x}^{\beta}_j)$ is a pair potential that describes the interaction between atom $\alpha$ of molecule $i$ and atom $\beta$ of molecule $j$. Finally, $\bold{f}^{\alpha}_i$ is an applied torque or an active driving force acting on atom $\alpha$ of molecule $i$, which is assumed to act in an equal and opposite manner on atoms 1 and 2 of molecule $i$, \emph{i.e.}, $\bold{f}^{1}_i = - \bold{f}^{2}_i = \bold{f}_i$. Here and in what follows, $\dot{\overline{( \cdot )}} = \dfrac{\mathrm{d}}{\mathrm{d} t}( \cdot )$ denotes the total time derivative.
\subsection{Microscopic-Macroscopic Relations}\label{sec:ik_relations}

To write the balance equations for mass, linear momentum, angular momentum, moment-of-intertia, and energy, the relationships between microscopic and macroscopic variables need to be defined. In doing so, we  follow the Irving-Kirkwood procedure \cite{irving1950statistical} by proposing relations between the extensive quantities. To this end, the mass density $\rho(\bold{x},t)$ at any macroscopic spatial point $\bold{x}$ and time $t$ is defined as 
\begin{equation}\label{eq:microscopic_mass}
\rho(\bold{x},t)=\sum_{i, \alpha}m^{\alpha}_i\Delta(\bold{x}-\bold{x}^{\alpha}_i) \ ,
\end{equation}
%
%
$\\$where $\Delta(\bold{x}-\bold{x}^{\alpha}_i)$ is a coarse-graining function that defines the contribution of atom $\alpha$ of molecule $i$ at the spatial point $\bold{x}$. Similarly, the linear momentum density $\rho(\bold{x},t) \bold{v}(\bold{x},t)$ is defined as 
\begin{equation}\label{eq:mom_rel}
\rho(\bold{x},t) \bold{v}(\bold{x},t) = \sum_{i, \alpha} \bold{p}^{\alpha}_i\Delta(\bold{x}-\bold{x}^{\alpha}_i) \ ,
\end{equation}
where $\bold{v}(\bold{x},t)$ is the velocity of the spatial point $\bold{x}$. The angular momentum density $\rho(\bold{x},t) \bold{L}(\bold{x},t)$ is defined to be 
\begin{equation}\label{eq:ang_mom_rel}
\rho(\bold{x},t) \bold{L}(\bold{x},t) =\sum_{i,\alpha} \bold{x}^{\alpha}_i \times \bold{p}^{\alpha}_i\Delta (\bold{x}-\bold{x}^{\alpha}_i) \ .
\end{equation} 
The moment-of-inertia $\bold{I}(\bold{x},t)$ is defined as
\begin{equation} \label{eq:inertia_tensor}
\begin{split}
\bold{I}(\bold{x},t) &= \sum_{i, \alpha} \left\{ m^{\alpha}_i \left[(\bold{x}^{\alpha}_i-\bold{x}) \cdot (\bold{x}^{\alpha}_i-\bold{x})\right] \bold{i} - m^{\alpha}_i (\bold{x}^{\alpha}_i -\bold{x}) \otimes (\bold{x}^{\alpha}_i-\bold{x}) \right\} \Delta(\bold{x}-\bold{x}^{\alpha}_i) \\
 &=\sum_{i, \alpha} \bold{I}^{\alpha}_i \Delta(\bold{x}-\bold{x}^{\alpha}_i) \ ,
 \end{split}
\end{equation}
where
\begin{equation}\label{eq:micro_inertia}
\bold{I}^{\alpha}_i =   m^{\alpha}_i \left[ (\bold{x}^{\alpha}_i-\bold{x}) \cdot (\bold{x}^{\alpha}_i-\bold{x})\right] \bold{i} - m^{\alpha}_i (\bold{x}^{\alpha}_i -\bold{x}) \otimes (\bold{x}^{\alpha}_i-\bold{x}) \ ,
\end{equation}
and $\bold{i}$ is the identity tensor. In \eqref{eq:inertia_tensor} and \eqref{eq:micro_inertia}, the symbol $``\otimes
"$ denotes the dyadic product or tensor outer product, where $\ab \otimes \bb$ for any two vectors $\ab$ and $\bb$ yields a second order tensor with elements $a_ib_j$ in Einstein's indicial notation. Finally, the total energy density $\rho(\bold{x},t) e(\bold{x},t)$ is defined to be
\begin{equation} \label{eq:total_energy}
\rho(\bold{x},t) e(\bold{x},t) = \sum_{i, \alpha} \left[ \frac{\bold{p}^{\alpha}_i \cdot \bold{p}^{\alpha}_i}{2 m^{\alpha}_i}  + \frac{1}{2} \sum_{j, \beta} u_2(\bold{x}^{\alpha}_i,\bold{x}^{\beta}_j)  \right] \Delta(\bold{x}-\bold{x}^{\alpha}_i)+\frac{1}{2}\sum_{i, \alpha} u^s(\bold{x}^1_i,\bold{x}^2_i) \Delta(\bold{x}-\bold{x}^{\alpha}_i) \ .
\end{equation}

The coarse graining function $\Delta(\bold{x}- \bold{x}^{\alpha}_i)$ depends on the scalar distance between the macroscopic point $\bold{x}$ and $\bold{x}^{\alpha}_i$ and satisfies the relation 
\begin{equation}\label{eq:cg_iden}
\frac{\partial \Delta(\bold{x}-\bold{x}^{\alpha}_i)}{\partial \bold{x}^{\alpha}_i} = -\frac{\partial \Delta(\bold{x}-\bold{x}^{\alpha}_i)}{\partial \bold{x}}  \ ;
\end{equation}
see \cite{mandadapu2012homogenization} for properties of the coarse-graining functions. 

\subsection{Balance of Mass}

In this section, we derive the macroscopic balance of mass using the relation \eqref{eq:microscopic_mass}. Taking the time derivative of \eqref{eq:microscopic_mass} yields 
\begin{equation}
\begin{split}\label{eq:mass_IK}
\dot{\rho}&=\sum_{i, \alpha} m^{\alpha}_i \left[\frac{\partial \Delta(\bold{x}-\bold{x}^{\alpha}_i)}{\partial \bold{x}} \cdot {\bold{v}} + \frac{\partial \Delta(\bold{x}-\bold{x}^{\alpha}_i)}{\partial \bold{x}^{\alpha}_i}\cdot \bold{v}^{\alpha}_i  \right] \\
&=\frac{\partial}{\partial \bold{x}} \left[ \sum_{i, \alpha} m^{\alpha}_i  \Delta(\bold{x}-\bold{x}^{\alpha}_i) \right]\cdot \bold{v} - \frac{\partial}{\partial \bold{x}}\cdot \left[ \sum_{i, \alpha} m^{\alpha}_i  \Delta(\bold{x}-\bold{x}^{\alpha}_i) \bold{v}^{\alpha}_i \right] \\
&=\frac{\partial \rho}{\partial \bold{x}}\cdot \bold{v} - \frac{\partial}{\partial \bold{x}} \cdot \left(\rho \bold{v} \right) 
\end{split}
\end{equation}
where the second equality in \eqref{eq:mass_IK} is obtained by using the identity \eqref{eq:cg_iden}, and the third equality is obtained by the relation \eqref{eq:mom_rel}. Note that the time derivative of the continuum position yields the continuum velocity, \emph{i.e.}, $\dot{\xb}  = \vb(\xb,t)$. Also, $``\dfrac{\partial}{\partial \xb} \cdot "$ in \eqref{eq:mass_IK} denotes the divergence operator. In \eqref{eq:mass_IK} and in what follows, we omit writing the explicit functional dependencies of the densities for clarity. Upon further simplification, equation \eqref{eq:mass_IK} reduces to the local form of balance of mass at the macroscopic level given by 
\begin{equation}\label{eq:mass_bal}
\dot{\rho} + \rho  \frac{\partial}{\partial \bold{x}} \cdot \bold{v} = 0 \ .
\end{equation}

\subsection{Balance of Linear Momentum}\label{sec:blm}

In this section, we derive the balance of linear momentum at the macroscopic level using the momentum relation \eqref{eq:mom_rel}. To this end, following the same procedure as for the balance of mass, taking the time derivate of both sides of equation \eqref{eq:mom_rel} and using the identity \eqref{eq:cg_iden} yields
\begin{equation} \label{eq:momentum_balance}
\begin{split}
\dot{\overline{\rho\bold{v}}}&=\sum_{i, \alpha}  \dot{\bold{p}}^{\alpha}_i \Delta(\bold{x}-\bold{x}^{\alpha}_i) 
+\sum_{i, \alpha} \bold{p}^{\alpha}_i \left[\frac{\partial \Delta(\bold{x}-\bold{x}^{\alpha}_i)}{\partial \bold{x}} \cdot \bold{v} \right]- 
\sum_{i, \alpha} \bold{p}^{\alpha}_i \left[\frac{\partial \Delta(\bold{x}-\bold{x}^{\alpha}_i)}{\partial \bold{x}} \cdot \frac{\bold{p}^{\alpha}_i}{m^{\alpha}_i}\right] \\
& = \sum_{i, \alpha}  \dot{\bold{p}}^{\alpha}_i \Delta(\bold{x}-\bold{x}^{\alpha}_i) + 
\left\{ \frac{\partial}{\partial \bold{x}} \left[ \sum_{i, \alpha} \bold{p}^{\alpha}_i \Delta(\bold{x}-\bold{x}^{\alpha}_i) \right] \right\} \bold{v}
-\frac{\partial}{\partial \bold{x}} \cdot \left[ \sum_{i, \alpha} \bold{p}^{\alpha}_i \otimes \frac{\bold{p}^{\alpha}_i}{m^{\alpha}_i} \Delta(\bold{x}-\bold{x}^{\alpha}_i) \right] \\
&= \sum_{i, \alpha}  \dot{\bold{p}}^{\alpha}_i \Delta(\bold{x}-\bold{x}^{\alpha}_i) + 
\left[\frac{\partial}{\partial \bold{x}}  (\rho \bold{v})\right] \bold{v}
-\frac{\partial}{\partial \bold{x}} \cdot \left[ \sum_{i, \alpha} m^{\alpha}_i \Big(\frac{\bold{p}^{\alpha}_i}{m^{\alpha}_i} - \bold{v}\Big) \otimes \Big(\frac{\bold{p}^{\alpha}_i}{m^{\alpha}_i}-\bold{v}\Big) \Delta(\bold{x}-\bold{x}^{\alpha}_i) \right] \\ 
& \hspace{4in}- \frac{\partial}{\partial \bold{x}} \cdot \Big( \rho \bold{v} \otimes \bold{v} \Big) \\
& = \sum_{i, \alpha}  \dot{\bold{p}}^{\alpha}_i \Delta(\bold{x}-\bold{x}^{\alpha}_i) - 
\rho \bold{v} \left(\frac{\partial}{\partial \bold{x}} \cdot \bold{v}\right) + \frac{\partial}{\partial \bold{x}} \cdot \bold{T}^\text{K} \ ,
\end{split}
\end{equation}
where the third equality is obtained by using the momentum relation \eqref{eq:mom_rel} and the tensor $\bold{T}^{\text{K}}$ in \eqref{eq:momentum_balance} is given by
\begin{equation} \label{eq:kinetic_stress}
\bold{T}^\text{K} = - \sum_{i, \alpha} m^{\alpha}_i \left(\frac{\bold{p}^{\alpha}_i}{m^{\alpha}_i} - \bold{v}\right) \otimes \left(\frac{\bold{p}^{\alpha}_i}{m^{\alpha}_i}-\bold{v}\right) \Delta(\bold{x}-\bold{x}^{\alpha}_i) \ .
\end{equation} 
Note that we have used the property $\dfrac{\partial}{\partial \xb} \cdot (\ab \otimes \bb)$ yields a vector with components $\dfrac{\partial}{\partial x_j} (a_i b_j)$ in Einstein's summation convention. Expanding the time derivative on the left hand side of \eqref{eq:momentum_balance} and using the balance of mass \eqref{eq:mass_bal} reduces \eqref{eq:momentum_balance} to 
\begin{equation}\label{eq:mom_bal_1}
\rho \dot{\bold{v}} =  \sum_{i, \alpha}  \dot{\bold{p}}^{\alpha}_i \Delta(\bold{x}-\bold{x}^{\alpha}_i) + \frac{\partial}{\partial \bold{x}} \cdot \bold{T}^\text{K} \ .
\end{equation}

Using the equations of motion \eqref{eq:motion}, the first term on the right hand side of \eqref{eq:momentum_balance} can be simplified as 
\begin{equation}
\begin{split} \label{eq:first_mom_rel}
\sum_{i, \alpha} \dot{\bold{p}}^{\alpha}_i \Delta(\bold{x}-\bold{x}^{\alpha}_i) &= \sum_{i, \alpha}\Big( -\zeta\frac{\bold{p}^{\alpha}_i}{m^{\alpha}_i}  +\sqrt{2k_BT\zeta}\frac{\mathrm{d}\bold{W}}{\mathrm{d}t} \Big) \Delta(\bold{x}-\bold{x}^{\alpha}_i) \\
&+ \sum_{i, \alpha} \left[  \sum_{j, \beta} \bold{F}^{\alpha \beta}_{ij}\Delta(\bold{x}-\bold{x}^{\alpha}_i)- \frac{\partial u_s(\bold{x}^{1}_i, \bold{x}^{2}_i )
}{\partial \bold{x}^{\alpha}_i}\Delta(\bold{x}-\bold{x}^{\alpha}_i) + \bold{f}^{\alpha}_i\Delta(\bold{x}-\bold{x}^{\alpha}_i)\right] \ .
\end{split}
\end{equation}
In \eqref{eq:first_mom_rel}, the term $\displaystyle \sum_{i, \alpha, j, \beta} \bold{F}^{\alpha \beta}_{ij} \Delta(\bold{x}-\bold{x}^{\alpha}_i)$ arises due to the forces between all the particles in the system and can be simplified as 
\begin{equation}
\begin{split} \label{eq:begin_virial}
\sum_{i ,\alpha, j, \beta} \bold{F}^{\alpha \beta}_{ij} \Delta(\bold{x}-\bold{x}^{\alpha}_i) =-\frac{1}{2}\sum_{i, \alpha, j, \beta} \left\{ \frac{\partial u_2(\bold{x}^{\alpha}_i,\bold{x}^{\beta}_j)}{\partial \bold{x}^{\alpha}_i} \left[ \Delta(\bold{x}-\bold{x}^{\alpha}_i)- \Delta(\bold{x}-\bold{x}^{\beta}_j) \right] \right\} \ ,
\end{split}
\end{equation}
where use is made of the relation
\begin{equation}\label{eq:central_force}
\dfrac{\partial u_2(\bold{x}^{\alpha}_i,\bold{x}^{\beta}_j)}{\partial \bold{x}^{\beta}_j}=-\dfrac{\partial u_2(\bold{x}^{\alpha}_i,\bold{x}^{\beta}_j)}{\partial \bold{x}^{\alpha}_i} \ ,
\end{equation}
as the pair potential $\displaystyle u_2(\bold{x}^{\alpha}_i,\bold{x}^{\beta}_j)$ depends only on the distances between the corresponding pair of atoms. Equation \eqref{eq:begin_virial} can be simplified using Noll's identity \cite{Noll_1955,Noll2_1955} given by 
\begin{equation}\label{eq:Noll}
\Delta(\bold{x}-\bold{x}^{\alpha}_i)- \Delta(\bold{x}-\bold{x}^{\beta}_j) = -\frac{\partial}{\partial \bold{x}} \cdot (\bold{x}^{\alpha \beta}_{ij}b^{\alpha \beta}_{ij}) \ ,
\end{equation}
where $\bold{x}^{\alpha \beta}_{ij} =\bold{x}^{\alpha}_i-\bold{x}^{\beta}_j$ and $b^{\alpha \beta}_{ij}$ is the bond function defined as 
\begin{equation}\label{eq:bond}
b^{\alpha \beta}_{ij} =\int_0^1 \mathrm{d} \lambda \ \Delta (\bold{x}-\lambda \bold{x}^{\alpha}_i+\bold{x}^{\alpha \beta}_{ij}) \ .
\end{equation}
Using \eqref{eq:Noll} and \eqref{eq:bond}, the right hand side of \eqref{eq:begin_virial} can be rewritten as
\begin{equation}\label{eq:manipulation_1_0}
\begin{split}
\sum_{i, \alpha, j, \beta} \bold{F}^{\alpha \beta}_{ij} \Delta(\bold{x}-\bold{x}^{\alpha}_i) = 
- \frac{1}{2}\sum_{i, j, \alpha, \beta} \bold{F}^{\alpha \beta}_{ij}\frac{\partial}{\partial \bold{x}} \cdot \left(\bold{x}^{\alpha \beta}_{ij}b^{\alpha \beta}_{ij} \right) & = -\frac{1}{2}\frac{\partial}{\partial \bold{x}} \cdot \left( \sum_{i, j, \alpha, \beta} \bold{F}^{\alpha \beta}_{ij} \otimes  \bold{x}^{\alpha \beta}_{ij}b^{\alpha \beta}_{ij}  \right) \\
&=\frac{1}{2}\frac{\partial}{\partial \bold{x}} \cdot\bold{T}^\text{V} \ ,
\end{split}
\end{equation}
where the tensor $\bold{T}^\text{V}$ is given by
\begin{equation} \label{eq:virial_stress}
\bold{T}^\text{V} =  - \frac{1}{2} \sum_{i, j, \alpha, \beta} \bold{F}^{\alpha \beta}_{ij} \otimes  \bold{x}^{\alpha \beta}_{ij}b^{\alpha \beta}_{ij} \ .
\end{equation}

Next, the term $\displaystyle \sum_{i, \alpha}\dfrac{\partial  u_s(\bold{x}^{1}_i, \bold{x}^{2}_i )}{\partial \bold{x}^{\alpha}_i} \Delta(\bold{x}-\bold{x}^{\alpha}_i)$ in \eqref{eq:first_mom_rel} can be rewritten as 
%
\begin{align}
\sum_{i, \alpha} -\dfrac{\partial  u_s(\bold{x}^{1}_i, \bold{x}^{2}_i )}{\partial \bold{x}^{\alpha}_i} \Delta(\bold{x}-\bold{x}^{\alpha}_i) 
&=\sum_{i} \left[ -\frac{\partial  u_s(\bold{x}^{1}_i, \bold{x}^{2}_i )}{\partial \bold{x}^1_i}\Delta(\bold{x}-\bold{x}^1_i)-\frac{\partial  u_s(\bold{x}^{1}_i, \bold{x}^{2}_i )}{\partial \bold{x}^2_i}\Delta(\bold{x}-\bold{x}^2_i)  \right] \nonumber \\
&=-\sum_{i} \frac{\partial  u_s(\bold{x}^{1}_i, \bold{x}^{2}_i )}{\partial \bold{x}^1_i} \left[ \Delta(\bold{x}-\bold{x}^1_i) -\Delta (\bold{x}-\bold{x}^2_i) \right]  \nonumber\\
&=\frac{\partial}{\partial \bold{x}} \cdot \left[ \sum_i \frac{\partial  u_s(\bold{x}^{1}_i, \bold{x}^{2}_i )}{\partial \bold{x}^1_i} \otimes \bold{x}^{12}_{ii} b^{12}_{ii} \right]  \nonumber\\
&=\frac{\partial}{\partial \bold{x}} \cdot \bold{T}^\text{S} \ ,
\label{eq:manipulation_1_1}
\end{align}
%
%
where $\bold{x}^{12}_{ii} = \bold{x}^{1}_{i} - \bold{x}^{2}_{i}$ and $\bold{T}^\text{S}$ is given by
\begin{equation}\label{eq:spring_stress}
\bold{T}^\text{S} = \sum_i \frac{\partial  u_s(\bold{x}^{1}_i, \bold{x}^{2}_i )}{\partial \bold{x}^1_i} \otimes \bold{x}^{12}_{ii} b^{12}_{ii} \ .
\end{equation}

Finally, the remaining term in (\ref{eq:first_mom_rel}) due to the active driving force can be simplified using the equal and opposite forces $\bold{f}_i$ acting on atoms $1$ and $2$ of molecule $i$ to obtain
\begin{equation}\label{eq:manipulation_1_2}
\begin{split}
\sum_{i, \alpha} \bold{f}^{\alpha}_i \Delta(\bold{x}-\bold{x}^{\alpha}_i) 
&=\sum_i \bold{f}_i \left[\Delta(\bold{x}-\bold{x}^1_i)-\Delta(\bold{x}-\bold{x}^2_i) \right]
=-\frac{\partial}{\partial \bold{x}} \cdot \left( \sum_i \bold{f}_i \otimes \bold{x}^{12}_{ii} b^{12}_{ii} \right) \\
 & = \frac{\partial}{\partial \bold{x}} \cdot \bold{T}^\text{A} \ ,
\end{split}
\end{equation}
where the tensor $\bold{T}^\text{A}$ is defined as
\begin{equation} \label{eq:active_stress}
\bold{T}^\text{A}= -  \sum_i \bold{f}_i \otimes \bold{x}^{12}_{ii} b^{12}_{ii}  \ .
\end{equation} 

Combining all the terms in \eqref{eq:manipulation_1_0}, \eqref{eq:manipulation_1_1}, and \eqref{eq:manipulation_1_2}, and substituting them into \eqref{eq:first_mom_rel} reduces \eqref{eq:first_mom_rel} to
\begin{equation}\label{eq:reduced_forces}
\begin{split}
\sum_{i, \alpha} \dot{\bold{p}}^{\alpha}_i \Delta(\bold{x}-\bold{x}^{\alpha}_i) &= \sum_{i, \alpha}\Big( -\zeta\frac{\bold{p}^{\alpha}_i}{m^{\alpha}_i}  +\sqrt{2k_BT\zeta}\frac{\mathrm{d}\bold{W}}{\mathrm{d}t} \Big) \Delta(\bold{x}-\bold{x}^{\alpha}_i)  + \frac{\partial}{\partial \bold{x}} \cdot (\bold{T}_\text{V} + \bold{T}_\text{S} + \bold{T}_\text{A} ) \ .
\end{split}
\end{equation}
Using \eqref{eq:reduced_forces}, equation \eqref{eq:mom_bal_1} can be reduced to 
\begin{equation}\label{eq:mom_bal_final}
\rho\dot{\bold{v}}=\frac{\partial}{\partial \bold{x}} \cdot \bold{T} +\rho\bold{b} \ ,
\end{equation}
where, $\bold{T}$ is the total stress tensor given by 
\begin{equation}\label{eq:sum_T}
\bold{T} = \left( \bold{T}^\text{K}+\bold{T}^\text{V}+\bold{T}^\text{A}+\bold{T}^\text{S} \right) \ ,
\end{equation}
and $\rho \bold{b}$ is the body force density given by 
\begin{equation}\label{eq:body_force}
\rho \bold{b} = \sum_{i, \alpha}\left( -\zeta\frac{\bold{p}^{\alpha}_i}{m^{\alpha}_i}  +\sqrt{2k_BT\zeta}\frac{\mathrm{d}\bold{W}}{\mathrm{d}t} \right) \Delta(\bold{x}-\bold{x}^{\alpha}_i) \ .
\end{equation}

Equation \eqref{eq:mom_bal_final} is a statement of the macroscopic balance of linear momentum with the total stress tensor $\bold{T}$ given by the sum of the stresses coming from kinetic terms $\bold{T}^\text{K}$, virial terms that include the interatomic interactions from the pair potentials $\bold{T}^\text{V}$, harmonic spring terms $\bold{T}^\text{S}$, and the active forces $\bold{T}^\text{A}$. It can be seen that the kinetic stresses $\bold{T}^\text{K}$, virial stresses $\bold{T}^\text{V}$, and harmonic stresses $\bold{T}^\text{S}$ are symmetric in nature. However, the active stress $\bold{T}^\text{A}$ is not symmetric. This asymmetry is very particular to a system consisting of microscopic couples as in the case of active dumbbells considered here. Note that the stress tensor can contain non-symmetric parts in the case of multi-body potentials \cite{Kranthi2009a, Kranthi_thesis}, for example, in the case of interactions consisting of three-body interactions \cite{Stillinger1985}, which are usually used to model systems such as water \cite{molinero2008water, limmer2011putative} and silicon \cite{Stillinger1985, KranthiSilicon}. Moreover, it should be noted that our work deviates from the analysis in \cite{hatwalne2004rheology}, where the stress tensor is considered for systems that contain contributions from active forces but still do not have an active rotational component that can lead to the asymmetry of the stress tensor. Specifically, in \cite{hatwalne2004rheology}, it is found that the active forces contribute to the deviatoric but still symmetric part of the stress tensor. This is also the case in \cite{lau2009fluctuating}, where contributions from the active forces still contain only symmetric terms. 

Given the nature of the stress tensor $\Tb$ in \eqref{eq:sum_T}, pressure in the system at any macroscopic point $\xb$ and time $t$ can be obtained as the negative of the trace of the stress tensor. As can be seen from the expressions of the individual components of the total stress tensor given by equations \eqref{eq:kinetic_stress}, \eqref{eq:virial_stress}, \eqref{eq:spring_stress}, and \eqref{eq:active_stress}, pressure consists of contributions only from kinetic and intermolecular forces, and not the active forces. This is due to the assumption in the microscopic dynamics \eqref{eq:motion} that the active forces act always perpendicular to the bond connecting the two atoms of the molecules thereby making $\Tb^{\text{A}}$ traceless. Hence, the active forces only contribute the deviatoric components of the stress tensor. 

%
\begin{figure}[!t]
	\centering
		\includegraphics[width=0.7\textwidth]{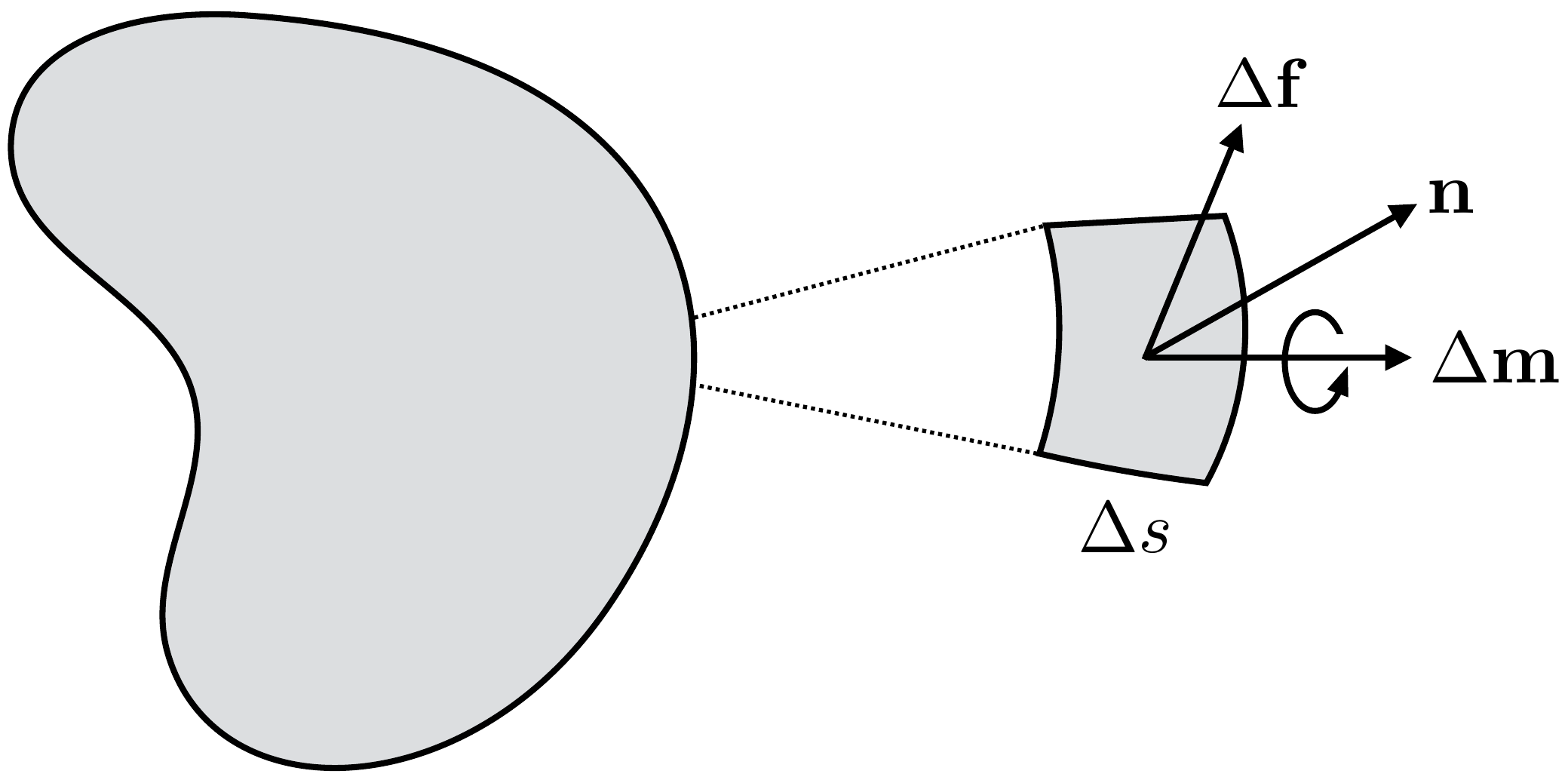}
	\captionsetup{width=0.80\textwidth}
	\caption{
		Stress and couples: A schematic showing the forces and couples acting on an infinitesimal area $\Delta s$ at a point on the surface of a body. Here, $\Delta \fb$ and $\Delta \mb$ are the surface forces and couples, related to the stress tensor and the couple stress tensor by the relations $\displaystyle \lim_{\Delta s \to 0 } \frac{\Delta \fb}{\Delta s} = \Tb \nb $, and $\displaystyle  \lim_{\Delta s \to 0 } \frac{\Delta \mb}{\Delta s} =  \Cb \nb$. 
	}
	\label{fig:couples}
\end{figure}
\subsection{Balance of Angular Momentum}
The asymmetric part of the stress tensor $\bold{T}$ drives a spin angular momentum, and understanding spin requires derivation of the balance of angular momentum. The balance of angular momentum has been proposed as a law at the continuum level by \cite{Dahler_1961} leading to the introduction of couple stress tensors based on the application of the surface couples; see Fig.~\ref{fig:couples}. In this section, we derive the balance of angular momentum from the microscopic dynamics given by \eqref{eq:motion} and identify the expressions for the couple stress tensor in terms of the microscopic variables.

The balance of angular momentum can be derived at the macroscopic level using the angular momentum relation \eqref{eq:ang_mom_rel} and following a procedure similar to the derivation of balance of linear momentum in Section~\ref{sec:blm}. To this end, taking the time derivate of both sides of \eqref{eq:ang_mom_rel} and employing the identity \eqref{eq:cg_iden} yields
\begin{equation} \label{eq:angular_momentum_balance}
\begin{split}
\dot{\overline{\rho\bold{L}}}&=\sum_{i, \alpha} \bold{x}^{\alpha}_i \times \dot{\bold{p}}^{\alpha}_i  \Delta(\bold{x}-\bold{x}^{\alpha}_i) 
+\sum_{i, \alpha} \bold{x}^{\alpha}_i \times \bold{p}^{\alpha}_i \left[\frac{\partial \Delta(\bold{x}-\bold{x}^{\alpha}_i)}{\partial \bold{x}} \cdot \bold{v} \right]- 
\sum_{i, \alpha} \bold{x}^{\alpha}_i \times \bold{p}^{\alpha}_i \Bigg(\frac{\partial \Delta(\bold{x}-\bold{x}^{\alpha}_i)}{\partial \bold{x}} \cdot \frac{\bold{p}^{\alpha}_i}{m^{\alpha}_i}\Bigg) \\
& = \sum_{i, \alpha} \bold{x}^{\alpha}_i \times \dot{\bold{p}}^{\alpha}_i  \Delta(\bold{x}-\bold{x}^{\alpha}_i) +  
\left\{ \frac{\partial}{\partial \bold{x}} \left[ \sum_{i,\alpha} \bold{x}^{\alpha}_i \times \bold{p}^{\alpha}_i \Delta(\bold{x}-\bold{x}^{\alpha}_i) \right] \right\} \bold{v} \\
&  \hspace{2.5in}
-\frac{\partial}{\partial \bold{x}} \cdot \bigg[ \sum_{i, \alpha} \big( \bold{x}^{\alpha}_i \times \bold{p}^{\alpha}_i \big) \otimes \frac{ \bold{p}^{\alpha}_i}{m^{\alpha}_i} \Delta(\bold{x}-\bold{x}^{\alpha}_i) \bigg]\\
&= \sum_{i, \alpha} \bold{x}^{\alpha}_i \times \dot{\bold{p}}^{\alpha}_i  \Delta(\bold{x}-\bold{x}^{\alpha}_i) + \bigg[\frac{\partial}{\partial \bold{x}} \big( \rho \bold{L} \big)\bigg]  \bold{v} \\
& \hspace{1.in}
- \frac{\partial}{\partial \bold{x}} \cdot \Big( \rho \bold{L} \otimes \bold{v} \Big)-
\frac{\partial}{\partial \bold{x}} \cdot \Bigg[ \sum_{i,\alpha} \Big( \bold{x}^{\alpha}_i \times \bold{p}^{\alpha}_i \Big) \otimes \Big( \frac{ \bold{p}^{\alpha}_i}{m^{\alpha}_i} -  \bold{v} \Big) \Delta(\bold{x}-\bold{x}^{\alpha}_i) \Bigg] \ ,
\end{split}
\end{equation}
where the third equality is obtained using the definition of macroscopic angular momentum \eqref{eq:ang_mom_rel}. Expanding the time derivative on the left hand side of \eqref{eq:angular_momentum_balance}, using the balance of mass \eqref{eq:mass_bal} and the following identity 
\begin{equation}
\frac{\partial}{\partial \bold{x}} \cdot \Big( \rho \bold{L} \otimes \bold{v} \Big)  =  \Big( \rho \frac{\partial}{\partial \bold{x}} \cdot \bold{v} \Big) \bold{L} + \bigg[\frac{\partial}{\partial \bold{x}} \big( \rho \bold{L} \big)\bigg]  \bold{v} \ ,
\end{equation}
equation \eqref{eq:angular_momentum_balance} can be reduced to 
\begin{equation} \label{eq:ang_mom_1}
\rho \dot{\Lb} = \sum_{i, \alpha} \bold{x}^{\alpha}_i \times \dot{\bold{p}}^{\alpha}_i  \Delta(\bold{x}-\bold{x}^{\alpha}_i) - \frac{\partial}{\partial \bold{x}} \cdot \Bigg[ \sum_{i,\alpha} \Big( \bold{x}^{\alpha}_i \times \bold{p}^{\alpha}_i \Big) \otimes \Big( \frac{ \bold{p}^{\alpha}_i}{m^{\alpha}_i} -  \bold{v} \Big) \Delta(\bold{x}-\bold{x}^{\alpha}_i) \Bigg] \ .
\end{equation}

The last term on the right hand side of \eqref{eq:ang_mom_1} can be rewritten as 
\begin{equation}\label{eq:am_sim_1}
\begin{split}
- \frac{\partial}{\partial \bold{x}} &\cdot \Bigg[ \sum_{i,\alpha} \Big( \bold{x}^{\alpha}_i \times \bold{p}^{\alpha}_i \Big)  \otimes \Big( \frac{ \bold{p}^{\alpha}_i}{m^{\alpha}_i} -  \bold{v} \Big) \Delta(\bold{x}-\bold{x}^{\alpha}_i) \Bigg] \\
&= - \frac{\partial}{\partial \bold{x}} \cdot \Bigg\{ \sum_{i,\alpha} \left[ (\bold{x}^{\alpha}_i - \xb) \times \bold{p}^{\alpha}_i \right]  \otimes \Big( \frac{ \bold{p}^{\alpha}_i}{m^{\alpha}_i} -  \bold{v} \Big) \Delta(\bold{x}-\bold{x}^{\alpha}_i) \Bigg\} \\ 
& \hspace{1.5in} - \frac{\partial}{\partial \bold{x}} \cdot \Bigg[ \sum_{i,\alpha} \Big( \bold{x} \times \bold{p}^{\alpha}_i \Big)  \otimes \Big( \frac{ \bold{p}^{\alpha}_i}{m^{\alpha}_i} -  \bold{v} \Big) \Delta(\bold{x}-\bold{x}^{\alpha}_i) \Bigg] \\
&=  \frac{\partial}{\partial \bold{x}} \cdot \Cb^\text{K}  - \frac{\partial}{\partial \bold{x}} \cdot \Bigg[ \sum_{i,\alpha} \Big( \bold{x} \times \bold{p}^{\alpha}_i \Big)  \otimes \Big( \frac{ \bold{p}^{\alpha}_i}{m^{\alpha}_i} -  \bold{v} \Big) \Delta(\bold{x}-\bold{x}^{\alpha}_i) \Bigg] \ ,
\end{split}
\end{equation}
where the tensor $\bold{C}^\text{K}$ is given by
\begin{equation}\label{eq:couple_stress_K}
\bold{C}^\text{K}=-\Bigg\{ \sum_{i,\alpha} \Big[ (\bold{x}^{\alpha}_i - \bold{x}) \times \bold{p}^{\alpha}_i \Big] \otimes \Big( \frac{\bold{p}^{\alpha}_i}{m^{\alpha}_i}-\bold{v} \Big)\Delta(\bold{x}-\bold{x}^{\alpha}_i) \Bigg\} \ .
\end{equation}
The second term on the right hand side of \eqref{eq:am_sim_1} can be simplified using Einstein's indicial notation techniques as 
\begin{equation}\label{eq:am_sim_2}
\begin{split}
\frac{\partial}{\partial \bold{x}} & \cdot \Bigg[ \sum_{i,\alpha} \Big( \bold{x} \times \bold{p}^{\alpha}_i \Big)  \otimes \Big( \frac{ \bold{p}^{\alpha}_i}{m^{\alpha}_i} -  \bold{v} \Big) \Delta(\bold{x}-\bold{x}^{\alpha}_i) \Bigg] \\
&= \frac{\partial}{\partial x_o} \left[ \sum_{i, \alpha} \epsilon_{lmn} x_m (\bold{p}^{\alpha}_i)_n \left( \frac{\bold{p}^{\alpha}_i}{m^{\alpha}_i} -\bold{v}  \right)_o \Delta(\bold{x}-\bold{x}^{\alpha}_i)  \right] \\
&=\sum_{i, \alpha} \epsilon_{lmn} \delta_{mo}(\bold{p}^{\alpha}_i)_n \left(\frac{\bold{p}^{\alpha}_i}{m^{\alpha}_i}-\bold{v}\right)_o \Delta(\bold{x}-\bold{x}^{\alpha}_i)+ \epsilon_{lmn} x_m \sum_{i, \alpha} \frac{\partial}{\partial x_o}\left[ (\bold{p}^{\alpha}_i)_n(\frac{\bold{p}^{\alpha}_i}{m^{\alpha}_i} -\bold{v})_o \Delta(\bold{x}-\bold{x}^{\alpha}_i) \right] \\
&=\sum_{i, \alpha} \epsilon_{lmn} m^{\alpha}_i \left( \frac{\bold{p}^{\alpha}_i}{m^{\alpha}_i} - \bold{v}  \right)_n \left( \frac{\bold{p}^{\alpha}_i}{m^{\alpha}_i} - \bold{v}  \right)_m\Delta(\bold{x}-\bold{x}^{\alpha}_i)+\sum_{i, \alpha} \epsilon_{lmn} m^{\alpha}_i v_n \left( \frac{\bold{p}^{\alpha}_i}{m^{\alpha}_i} -\bold{v} \right)_m\Delta(\bold{x}-\bold{x}^{\alpha}_i) \\
& \hspace{1.0in}+ \epsilon_{lmn} x_m \frac{\partial}{\partial x_o} \left[ \sum_{i, \alpha} m^{\alpha}_i \left( \frac{\bold{p}^{\alpha}_i}{m^{\alpha}_i} - \bold{v} \right)_n \left(\frac{p^{\alpha}_i}{m^{\alpha}_i} -\bold{v} \right)_o \Delta(\bold{x}-\bold{x}^{\alpha}_i) \right] \\
& \hspace{2.0in}+ \epsilon_{lmn} x_m  \frac{\partial}{\partial x_o}\left[  \sum_{i, \alpha}  m^{\alpha}_i v_n \left(\frac{\bold{p}^{\alpha}_i}{m^{\alpha}_i}-\bold{v} \right)_o\Delta(\bold{x}-\bold{x}^{\alpha}_i) \right]
\end{split}
\end{equation}
where $\epsilon_{lmn}$ is the permutation tensor and $\delta_{mn}$ is the Kronecker Delta. Noting that
\begin{equation}\label{eq:am_sim_3}
\sum_{i, \alpha} m^{\alpha}_i v_n \left( \frac{\bold{p}^{\alpha}_i}{m^{\alpha}_i} -\bold{v}\right)_o  \Delta(\bold{x}-\bold{x}^{\alpha}_i) = 0  \ , 
\end{equation}
due to the linear momentum relation \eqref{eq:mom_rel}, equation \eqref{eq:am_sim_2} can be reduced to
\begin{equation}\label{eq:am_sim_4}
\begin{split}
\frac{\partial}{\partial \bold{x}} \cdot \Bigg[ \sum_{i,\alpha} \Big( \bold{x} \times \bold{p}^{\alpha}_i \Big)  \otimes \Big( \frac{ \bold{p}^{\alpha}_i}{m^{\alpha}_i} -  \bold{v} \Big) \Delta(\bold{x}-\bold{x}^{\alpha}_i) \Bigg] & = -\epsilon_{lmn} T^{\text{K}}_{nm} - \epsilon_{lmn} x_m \frac{\partial}{\partial x_o}\left(T^{\text{K}}_{no}\right) \\ 
& = -\bold{A}^{\text{K}} - \bold{x} \times \frac{\partial}{\partial \bold{x}} \cdot \bold{T}^{\text{K}} 
\end{split}
\end{equation}
where ${T}^\text{K}_{mn}$ are the components of the kinetic part of the stress tensor $\bold{T}^\text{K}$ in the indicial notation, $\bold{A}^{\text{K}}$ denotes the vector formed from the anti-symmetric part of the kinetic stress tensor with its components $A^{\text{K}}_l = \epsilon_{lmn} T^{\text{K}}_{nm}$. Using equations \eqref{eq:am_sim_1}, \eqref{eq:couple_stress_K} and \eqref{eq:am_sim_4}, equation \eqref{eq:ang_mom_1} can be simplified to 
\begin{equation} \label{eq:ang_mom_2}
\rho \dot{\Lb} = \sum_{i, \alpha} \bold{x}^{\alpha}_i \times \dot{\bold{p}}^{\alpha}_i  \Delta(\bold{x}-\bold{x}^{\alpha}_i) +  \frac{\partial}{\partial \bold{x}} \cdot \Cb^\text{K} + \bold{A}^{\text{K}} + \bold{x} \times \frac{\partial}{\partial \bold{x}} \cdot \bold{T}^{\text{K}}  \ .
\end{equation}

The remaining term that can be reduced is the first term on the right hand side of equation \eqref{eq:ang_mom_2}, which upon using the microscopic equations of motion \eqref{eq:motion} yields, 
\begin{equation}\label{eq:am_split_1}
\begin{split}
\sum_{i, \alpha} \bold{x}^{\alpha}_i \times \dot{\bold{p}}^{\alpha}_i\Delta(\bold{x}-\bold{x}^{\alpha}_i)  = \sum_{i, \alpha} \bold{x}^{\alpha}_i \times \left(- \zeta \frac{\bold{p} ^{\alpha} _i}{m_i} +  \sum_{j, \beta} \bold{F}^{\alpha \beta}_{ij} +\bold{f}^{\alpha}_i-\frac{\partial u^s(\bold{x}^1_i,\bold{x}^2_i)}{\partial \bold{x}^{\alpha}_i} +\sqrt{2k_B T\zeta}\frac{d\bold{W}}{dt}  \right) \ .
\end{split}
\end{equation}

In what follows, we manipulate the terms on the right hand side of \eqref{eq:am_split_1} to derive the quantities in the balance of angular momentum at the coarse-grained level. To this end, the second term on the right hand side of \eqref{eq:am_split_1} can be manipulated to yield 
\begin{equation}\label{eq:am_split_1_1}
\begin{split}
\sum_{i, \alpha}& \bold{x}^{\alpha}_i \times  \sum_{j, \beta} \bold{F}^{\alpha \beta}_{ij} \\
& = \sum_{i, \alpha, j, \beta} \bold{x}^{\alpha}_i \times \frac{-\partial u_2 (\bold{x}^{\alpha}_i,\bold{x}^{\beta}_j)}{\partial \bold{x}^{\alpha}_i}\Delta(\bold{x}-\bold{x}^{\alpha}_i) \\
&=-\frac{1}{2} \sum_{i, \alpha, j, \beta} \left( \bold{x}^{\alpha}_i \times \frac{\partial u_2 (\bold{x}^{\alpha}_i,\bold{x}^{\beta}_j)}{\partial \bold{x}^{\alpha}_i}\Delta(\bold{x}-\bold{x}^{\alpha}_i) + \bold{x}^{\beta}_j \times \frac{ \partial u_2 (\bold{x}^{\alpha}_i,\bold{x}^{\beta}_j)}{\partial \bold{x}^{\beta}_j}\Delta(\bold{x}-\bold{x}^{\beta}_j) \right) \\
&=-\frac{1}{2}\sum_{i, \alpha, j, \beta} \left[ \bold{x}^{\alpha}_i \Delta(\bold{x}-\bold{x}^{\alpha}_i) - \bold{x}^{\beta}_j\Delta(\bold{x}-\bold{x}^{\beta}_j)   \right] \times \frac{\partial u_2 (\bold{x}^{\alpha}_i,\bold{x}^{\beta}_j)}{\partial \bold{x}^{\alpha}_i} \\
&=-\frac{1}{2}\sum_{i, \alpha, j, \beta} \Bigg[ \bold{x}^{\alpha}_i \Delta(\bold{x}-\bold{x}^{\alpha}_i) - \bold{x}^{\alpha}_i \Delta(\bold{x}-\bold{x}^{\beta}_j) + \bold{x}^{\alpha}_i \Delta(\bold{x}-\bold{x}^{\beta}_j)  - \bold{x}^{\beta}_j\Delta(\bold{x}-\bold{x}^{\beta}_j)   \Bigg] \times \frac{\partial u_2 (\bold{x}^{\alpha}_i,\bold{x}^{\beta}_j)}{\partial \bold{x}^{\alpha}_i} \\
&=\frac{1}{2}\sum_{i, \alpha, j, \beta} \left[ -\bold{x}^{\alpha}_i \left( \frac{\partial}{\partial \bold{x}} \cdot (\bold{x}^{\alpha \beta}_{ij} b^{\alpha \beta}_{ij})\right) + \bold{x}^{\alpha \beta}_{ij} \Delta(\bold{x}-\bold{x}^{\beta}_j)  \right] \times \bold{F}^{\alpha \beta}_{ij} \\
&=-\frac{1}{2}\frac{\partial}{\partial \bold{x}} \cdot \left[ \sum_{i, \alpha, j, \beta} (\bold{x}^{\alpha}_i \times \bold{F}^{\alpha \beta}_{ij}) \otimes \bold{x}^{\alpha \beta}_{ij} b^{\alpha \beta}_{ij}  \right]+\frac{1}{2}\sum_{i, \alpha, j, \beta}\bold{x}^{\alpha \beta}_{ij} \times \bold{F}^{\alpha \beta}_{ij}\Delta(\bold{x}-\bold{x}^{\beta}_j) \\
&=\frac{\partial}{\partial \bold{x}} \cdot \left[ -\frac{1}{2}\sum_{i, \alpha, j, \beta} (\bold{x}^{\alpha}_i -\bold{x}) \times \bold{F}^{\alpha \beta}_{ij} \otimes \bold{x}^{\alpha \beta}_{ij} b^{\alpha \beta}_{ij}  \right] \\
& \hspace{0.25in} + \frac{\partial}{\partial \bold{x}} \cdot \left[  -\frac{1}{2}\sum_{i, \alpha, j, \beta} \left(\bold{x} \times \bold{F}^{\alpha \beta}_{ij} \right) \otimes \bold{x}^{\alpha \beta}_{ij} b^{\alpha \beta}_{ij}  \right]+\frac{1}{2}\sum_{i, \alpha, j, \beta}\bold{x}^{\alpha \beta}_{ij} \times \bold{F}^{\alpha \beta}_{ij}\Delta(\bold{x}-\bold{x}^{\beta}_j) \\
&=\frac{\partial}{\partial \bold{x}} \cdot \bold{C}^{\text{V}}  + \epsilon_{lmn}T^{\text{V}}_{nm} + \bold{x} \times \frac{\partial}{\partial \bold{x}} \cdot \bold{T}^{\text{V}}+ \frac{1}{2}\sum_{i, \alpha, j, \beta}\bold{x}^{\alpha \beta}_{ij} \times \bold{F}^{\alpha \beta}_{ij}\Delta(\bold{x}-\bold{x}^{\beta}_j) \\ 
&=\frac{\partial}{\partial \bold{x}} \cdot \bold{C}^{\text{V}}  + \bold{A}^{\text{V}} + \bold{x} \times \frac{\partial}{\partial \bold{x}} \cdot \bold{T}^{\text{V}}+ \frac{1}{2}\sum_{i, \alpha, j, \beta}\bold{x}^{\alpha \beta}_{ij} \times \bold{F}^{\alpha \beta}_{ij}\Delta(\bold{x}-\bold{x}^{\beta}_j) \ , 
\end{split}
\end{equation}
where the tensor $\bold{C}^{\text{V}}$ is given by 
\begin{equation}\label{eq:couple_stress_V}
\bold{C}^{\text{V}}= -\frac{1}{2}\left( \sum_{i, \alpha , j , \beta} (\bold{x}^{\alpha}_i -\bold{x}) \times \bold{F}^{\alpha \beta}_{ij} \otimes \bold{x}^{\alpha \beta}_{ij} b^{\alpha \beta}_{ij}  \right) \ ,
\end{equation}
$\bold{A}^{\text{V}}$ is the vector formed by the antisymmetric parts of the virial stress tensor $\bold{T}^{\text{V}}$ given by $A^{\text{V}}_l =  + \epsilon_{lmn}T^{\text{V}}_{nm} $, and the last equality in \eqref{eq:am_split_1_1} is obtained by using the pair potential contribution to the total stress tensor given in \eqref{eq:virial_stress}. The last term on the right hand side of \eqref{eq:am_split_1_1} is identically zero for systems modeled by pair potentials due to the fact that the force $\bold{F}^{\alpha \beta}_{ij} \propto \bold{x}^{\alpha \beta}_{ij}$. 

Using similar procedure to reduce the pair-potential mediated forces in \eqref{eq:am_split_1_1} and \eqref{eq:couple_stress_V}, the fourth term on the right-hand side of \eqref{eq:am_split_1} can be reduced to 
\begin{equation}\label{eq:am_split_1_2}
\begin{split}
\sum_{i, \alpha} \bold{x}^{\alpha}_i \times -\frac{\partial u^s(\bold{x}^1_i,\bold{x}^2_i)}{\partial \bold{x}^{\alpha}_i}  & =  \sum_i \bigg[ \bold{x}^1_i \times \left(-\frac{\partial u^s(x^{12}_i)}{\partial \bold{x}^1_i} \right) \Delta(\bold{x}-\bold{x}^{1}_i) + \bold{x}^2_i \times \left(-\frac{\partial u^s(x^{12}_i }{\partial \bold{x}^2_i}\right)\Delta(\bold{x}-\bold{x}^{2}_i)  \bigg] \\
&=\frac{\partial}{\partial \bold{x}} \cdot \bold{C}^{\text{S}}+  \bold{A}^{\text{S}} + \bold{x} \times \frac{\partial}{\partial \bold{x}} \cdot \bold{T}^{\text{S}}\ ,
\end{split}
\end{equation}
where the tensor $\bold{C}^{\text{S}}$ is given by 
\begin{equation}\label{eq:couple_stress_S}
\bold{C}^{\text{S}}=-\frac{1}{2} \sum_i(\bold{x}^1_i-\bold{x}) \times \frac{- \partial  u_s(\bold{x}^{1}_i, \bold{x}^{2}_i )}{\partial \bold{x}^1_i} \otimes \bold{x}^{12}_{ii} b^{12}_{ii} \ ,
\end{equation}
and $\bold{A}^{\text{S}}$ is the vector formed from the anti-symmetric part of the stress tensor $\bold{T}^{\text{S}}$.

Next, the third term on the right hand side of \eqref{eq:am_split_1_1} involving the active forces can be reduced to 
\begin{equation}\label{eq:am_split_1_3}
\begin{split}
 \sum_{i, \alpha} \bold{x}^{\alpha}_i \times \bold{f}^{\alpha}_i &=\sum_{i}  \left[ \bold{x}^1_i \Delta(\bold{x}-\bold{x}^{1}_i) -  \bold{x}^2_i \Delta(\bold{x}-\bold{x}^{2}_i) \right] \times \bold{f}_i  \\
&=\sum_i \left[ \bold{x}^1_i \Delta(\bold{x}-\bold{x}^{1}_i)  -\bold{x}^1_i \Delta(\bold{x}-\bold{x}^{2}_i) +\bold{x}^1_i \Delta(\bold{x}-\bold{x}^{2}_i) - \bold{x}^2_i \Delta(\bold{x}-\bold{x}^{2}_i) \right] \times \bold{f}_i \\
&=\sum_i \bold{x}^1_i \left[ -\frac{\partial}{\partial \bold{x}} \cdot (\bold{x}^{12}_{ii} b^{12}_{ii}) \right] \times \bold{f}_i + \sum_i \bold{x}^{12}_{ii} \times \bold{f}_i \Delta(\bold{x}-\bold{x}^{2}_i) \\
&=\frac{\partial}{\partial \bold{x}} \cdot \left[ - \sum_i (\bold{x}^1_i \times \bold{f}_i) \otimes \bold{x}^{12}_{ii} b^{12}_{ii}  \right]+\sum_i \bold{x}^{12}_{ii} \times \bold{f}_i \Delta(\bold{x}-\bold{x}^{2}_i) \\
&=\frac{\partial}{\partial \bold{x}} \cdot \left[ - \sum_i (\bold{x}^1_i -\bold{x}) \times \bold{f}_i \otimes \bold{x}^{12}_{ii} b^{12}_{ii}  \right]-\frac{\partial}{\partial \bold{x}} \cdot \left[ \sum_i (\bold{x} \times \bold{f}_i )\otimes \bold{x}^{12}_{ii} b^{12}_{ii}  \right] \\ 
& \hspace{2in} +\sum_i \bold{x}^{12}_{ii} \times \bold{f}_i \Delta(\bold{x}-\bold{x}^{2}_i) \\
&=\frac{\partial}{\partial \bold{x}} \cdot \bold{C}^{\text{A}}+ \bold{A}^{\text{A}} + \bold{x} \times \frac{\partial}{\partial \bold{x}} \cdot \bold{T}^{\text{A}} + \sum_i \bold{x}^{12}_{ii} \times \bold{f}_i \Delta(\bold{x}-\bold{x}^{2}_i) \ ,
\end{split}
\end{equation}
where the tensor $\bold{C}^{\text{A}}$ is defined as 
\begin{equation}\label{eq:couple_stress_A}
\bold{C}^{\text{A}}=- \sum_i [(\bold{x}^1_i -\bold{x}) \times \bold{f}_i ]\otimes \bold{x}^{12}_{ii} b^{12}_{ii}  \ , 
\end{equation}
and $\bold{A}^{\text{A}}$ is the vector formed by the anti-symmetric components of the active part of the stress tensor $\bold{T}^{\text{A}}$ similar to $\bold{A}^{\text{K}}$ and $\bold{A}^{\text{V}}$.

Finally, combining all the terms in equations \eqref{eq:am_split_1_1}-\eqref{eq:am_split_1_3}, the resulting angular momentum balance given by \eqref{eq:ang_mom_2} can be reduced to
\begin{equation}\label{eq:am_split_4}
\begin{split}
\sum_{i, \alpha} \bold{x}^{\alpha}_i \times \dot{\bold{p}}^{\alpha}_i\Delta(\bold{x}-\bold{x}^{\alpha}_i)   & = \sum_{i, \alpha} \bold{x}^{\alpha}_i \times \left(- \zeta \frac{\bold{p}^{\alpha} _i}{m_i} +\sqrt{2k_B T\zeta}\frac{d\bold{W}}{dt}  \right) + \sum_i \bold{x}^{12}_{ii} \times \bold{f}_i \Delta(\bold{x}-\bold{x}^{2}_i) + \\ 
& \hspace{0.4in} \frac{\partial}{\partial \bold{x}} \cdot \left( \bold{C}^{\text{K}} +  \bold{C}^{\text{V}}+  \bold{C}^{\text{S}} +  \bold{C}^{\text{S}}  \right) +  \bold{A}^{\text{K}} +  \bold{A}^{\text{S}} +  \bold{A}^{\text{V}} +  \bold{A}^{\text{A}}  \\
& \hspace{1.5in} + \bold{x} \times   \frac{\partial}{\partial \bold{x}} \cdot \left( \bold{T}^{\text{K}} + \bold{T}^{\text{V}} + \bold{T}^{\text{S}} + \bold{T}^{\text{A}} \right)  \\ 
& = \frac{\partial}{\partial \bold{x}} \cdot  \bold{C}  + \rho \bold{G} + \bold{x} \times \rho \bold{b} + 
\bold{A}+ \bold{x} \times \frac{\partial}{\partial \bold{x}} \cdot \bold{T} \ ,
\end{split}
\end{equation}
where 
\begin{equation}\label{eq:body_torque}
\rho \bold{G} = \sum_{i , \alpha}\bold{x}^{12}_{ii} \times \bold{f}_i \, \Delta(\bold{x}-\bold{x}^{2}_i)+\sum_{i , \alpha} \left( \bold{x}^{\alpha}_i -\bold{x} \right) \times \left( -\zeta  \frac{\bold{p}^{\alpha} _i}{m_i}+\sqrt{2 k_B T \zeta}\frac{d\bold{W}}{dt} \right)\Delta(\bold{x}-\bold{x}^{\alpha}_i)  \ ,
\end{equation} 
\begin{equation}\label{eq:C_sum}
\bold{C} =  \bold{C}^{\text{K}} +  \bold{C}^{\text{V}}+  \bold{C}^{\text{S}} +  \bold{C}^{\text{A}}   \ ,
\end{equation} 
and 
\begin{equation}\label{eq:A_sum}
\bold{A} =  \bold{A}^{\text{K}} +  \bold{A}^{\text{V}} +  \bold{A}^{\text{S}} +  \bold{A}^{\text{A}} \ .
\end{equation} 

Substituting \eqref{eq:am_split_4} in \eqref{eq:ang_mom_2}, the total balance of angular momentum at the coarse-grained level can be obtained as 
\begin{equation}\label{eq:ang_mom_bal_final}
\rho \dot{\bold{L}}= \frac{\partial}{\partial \bold{x}} \cdot  \bold{C}  + \rho \bold{G} + \bold{x} \times \rho \bold{b} + 
\bold{A}+ \bold{x} \times \frac{\partial}{\partial \bold{x}} \cdot \bold{T} \ . 
\end{equation}

Equation \eqref{eq:ang_mom_bal_final} is a statement of the macroscopic balance of linear momentum with the total couple-stress tensor $\bold{C}$ given by the sum of a contribution of the couple-stresses coming from kinetic $\bold{C}^\text{K}$, interatomic potential $\bold{C}^\text{V}$, harmonic spring $\bold{C}^\text{S}$, and active forces $\bold{C}^\text{A}$ and $\rho \bold{G}$ being the body torque at the coarse-grained level. As can be seen from the form in \eqref{eq:couple_stress_V}, the couple-stress is the virial contribution from the torque created by a force between the bond connecting the atoms $\{i,\alpha\}$ and $\{j,\beta\}$ with respect to the center $\bold{x}$ of the coarse-graining volume. The physical meaning of all the other terms $\bold{C}^\text{K}, \bold{C}^\text{S}$, and $\bold{C}^\text{A}$ can be understood in a similar way to $\bold{C}^\text{V}$. 

The microscopic derivation of balance of angular momentum has not been considered before by Irving and Kirkwood.  It is of interest to see that even though the stress tensor from interaction potentials and kinetic terms lead to a symmetric form for the stress tensor, the corresponding couple-stresses are not zero, which are usually neglected or not considered in the case of mechanics of continuous media, even in the case of passive particles \cite{landau1986theory}. This assumption is satisfied when the time scales associated with the relaxation of couple stresses are small compared to that of the stress tensor in addition to no body torques \cite{deGroot_1984}. In this case, the change of angular momentum and couple stresses can be ignored and the balance of angular momentum \eqref{eq:ang_mom_bal_final} then imposes the symmetry of the stress tensor. However, should the active forces driving the dumbbell particles be non-zero, the active couple-stresses $\bold{C}^{\text{A}}$ and the asymmetric part of the stress tensor from $\bold{T}^{\text{A}}$ need not be negligible. Moreover, the non-zero active forces also drive the body torques $\rho \Gb$ as can be seen from \eqref{eq:body_torque}. Therefore, the case of active dumbbells or any system consisting of microscopically rotating particles, presents a unique case where the spin stresses and the non-symmetric part of the stress tensor are not negligible, and the coupling between the linear and angular momentum should be considered. This can be seen explicitly from the balance of spin that is considered in the next section.

\subsubsection{Balance of Spin}

In this section, we introduce the concept of spin and derive the balance of spin. Following the work of Dahler and Scriven \cite{Dahler_1961}, the total angular momentum can be divided without loss of generality as   
\begin{equation} \label{total_angular}
\rho\bold{L} = \rho\bold{x} \times \bold{v} + \rho \bold{M} \ ,
\end{equation}
where $\rho \bold{x} \times \bold{v}$ is moment of linear momentum (or orbital angular momentum) and $\bold{M}$ is the internal spin, as shown in figure~\ref{fig2}.
\begin{figure}[!th]
	\centering
	\includegraphics[width=0.6\linewidth]{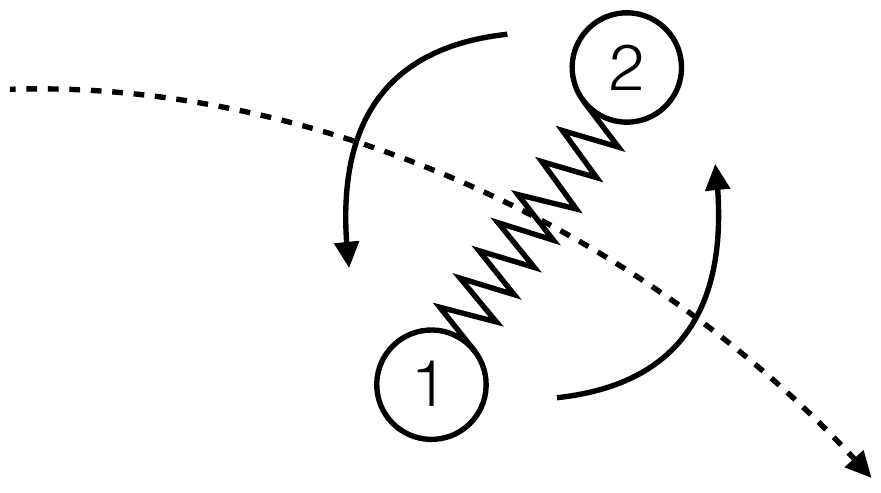}
	\captionsetup{width=0.80\textwidth}
	\caption{\label{fig2} A schematic showing the decomposition of the total angular momentum into orbital angular momentum and spin.}
\end{figure}

Taking the cross product of the local form of the balance of linear momentum with $\bold{x}$ yields, 
\begin{equation}\label{eq:cross_mom}
\bold{x} \times \rho \dot{\bold{v}} = \bold{x} \times  \frac{\partial}{\partial \bold{x}} \cdot \bold{T}+\rho \bold{x} \times \bold{b} \ .
\end{equation}
Making use of \eqref{total_angular} and \eqref{eq:cross_mom} in the local form of total angular momentum \eqref{eq:ang_mom_bal_final} yields the equation for balance of spin as 
\begin{equation}
\label{eq:final_spin}
\rho \dot{\bold{M}}=\frac{\partial}{\partial \bold{x}} \cdot \bold{C} + \rho \bold{G} +\bold{A} \ . 
\end{equation}
%

\subsection{Balance of Moment of Inertia}
Since the particles in a fluid are continuously exchanged in the neighborhood of the macroscopic point $\xb$, the moment of inertia $\Ib$ is transported across the system requiring the balance equation for moment of inertia. To this end, we derive the associated macroscopic balance of moment of inertia using the relation \eqref{eq:inertia_tensor}. Taking the time derivative of both sides of \eqref{eq:inertia_tensor} yields
 \begin{equation} \label{eq:moment_of_inertia_balance}
 \begin{split}
  \dot{\bold{I}} &= 2 \sum_{i,\alpha} \left[m^{\alpha}_i \left( \frac{\bold{p}^{\alpha}_i}{m^{\alpha}_i} -\bold{v} \right) \cdot \left( \bold{x}^{\alpha}_i -\bold{x} \right)\right] \bold{i} \  \Delta(\bold{x}-\bold{x}^{\alpha}_i) \\
  &  
 \hspace{28pt} - \sum_{i,\alpha}  \Bigg[ \left( \frac{\bold{p}^{\alpha}_i}{m^{\alpha}_i} - \bold{v} \right) \otimes \left( \bold{x}^{\alpha}_i -\bold{x} \right) + \left( \bold{x}^{\alpha}_i - \bold{x} \right) \otimes \left( \frac{\bold{p}^{\alpha}_i}{m^{\alpha}_i} - \bold{v} \right)  \Bigg] m^{\alpha}_i \Delta(\bold{x}-\bold{x}^{\alpha}_i) \\
 & \hspace{56pt} +\sum_{i,\alpha} \bold{I}^{\alpha}_i \left[ \frac{\partial \Delta(\bold{x}-\bold{x}^{\alpha}_i)}{\partial \bold{x}} \cdot \bold{v} - \frac{\partial \Delta(\bold{x}-\bold{x}^{\alpha}_i)}{\partial \bold{x}} \cdot \frac{\bold{p}^{\alpha}_i}{m^{\alpha}_i} \right] \ ,
 \end{split}
 \end{equation}
where we have used the identity \eqref{eq:cg_iden}. The last term on the right can be simplified as 
\begin{equation} \label{eq:mom_of_inertia_1}
\begin{split} 
&\sum_{i,\alpha} \bold{I}^{\alpha}_i \left[ \frac{\partial \Delta(\bold{x}-\bold{x}^{\alpha}_i)}{\partial \bold{x}} \cdot \bold{v} - \frac{\partial \Delta(\bold{x}-\bold{x}^{\alpha}_i)}{\partial \bold{x}} \cdot \frac{\bold{p}^{\alpha}_i}{m^{\alpha}_i} \right] \\
& \hspace{10pt}  = \frac{\partial}{\partial \bold{x}} \cdot \left[ \sum_{i, \alpha} \bold{I}^{\alpha}_i \otimes \left(\bold{v}-\frac{\bold{p}^{\alpha}_i}{m^{\alpha}_i}\right) \Delta(\bold{x}-\bold{x}^{\alpha}_i) \right] - \bold{I} \left( \frac{\partial}{\partial \bold{x}} \cdot \bold{v} \right)-2 \sum_{i, \alpha} \left[m^{\alpha}_i (\bold{x}-\bold{x}^{\alpha}_i) \cdot \left(\bold{v}-\frac{\bold{p}^{\alpha}_i}{m^{\alpha}_i} \right)\right] \bold{i} \\
& \hspace{56pt} - \sum_{i ,\alpha} m^{\alpha}_i \left[ (\bold{x}-\bold{x}^{\alpha}_i) \otimes \left(\frac{\bold{p}^{\alpha}_i}{m^{\alpha}_i} - \bold{v} \right) + \left(\frac{\bold{p}^{\alpha}_i}{m^{\alpha}_i} - \bold{v} \right) \otimes (\bold{x}-\bold{x}^{\alpha}_i) \right]\Delta(\bold{x}-\bold{x}^{\alpha}_i)  \ ,
\end{split}
\end{equation}
where the dyadic product is between a tensor and a vector, which for any general tensor $\Sb$ and vector $\db$ yields a third order tensor with the elements $S_{ij} b_k$ in the indicial notation.
Using \eqref{eq:mom_of_inertia_1}, the rate of change of moment of inertia in \eqref{eq:moment_of_inertia_balance} can be reduced to 
 \begin{equation} \label{eq:moi_bal_final}
 \dot{\bold{I}} +\bold{I} \left( \frac{\partial}{\partial \bold{x}} \cdot \bold{v} \right) = - \frac{\partial}{\partial \bold{x}} \cdot \Yb \ ,
 \end{equation}
where $\Yb$ is the moment of inertia flux tensor given 
\begin{equation}
\Yb = - \left[ \sum_{i, \alpha} \bold{I}^{\alpha}_i \otimes \left( \frac{\bold{p}^{\alpha}_i}{m^\alpha_i} - \bold{v} \right) \Delta(\bold{x}-\bold{x}^{\alpha}_i) \right] \ .
\end{equation}
Note that the moment of inertia flux tensor $\Yb$ is a third order tensor owing to the second order nature of the tensor $\Ib$. Equation \eqref{eq:moi_bal_final} is a statement of the macroscopic balance of momentum of inertia of the system.

Finally, without loss of any generality, the spin can be rewritten as 
\begin{equation}\label{eq:def_omega}
\rho \bold{M} = \bold{I}\boldsymbol{\omega}(\bold{x},t)  \ ,
\end{equation}
where $\omegab$ is the rotational velocity of the macroscopic point. Note that the definition of the moment of inertia $\Ib$, and thus $\omegab$, depends on the choice of the coordinate system. Using \eqref{eq:def_omega}, \eqref{eq:moi_bal_final} and \eqref{eq:mass_bal}, the balance of spin in \eqref{eq:final_spin} can be written in terms of the rate of rotation as 
\begin{equation}\label{eq:moi_bal_rot}
\Ib \dot{\omegab} = \left( \frac{\partial}{\partial \bold{x}} \cdot \Yb \right) \omegab + \frac{\partial}{\partial \bold{x}} \cdot \bold{C} + \rho \bold{G} +\bold{A} \ . 
\end{equation}

\subsection{Balance of Energy}\label{sec:ik_energy}
In this section, we derive the balance of total energy to identify the sources of heat and heat flux at the macroscopic scale. We begin by decomposing the total energy into the internal energy, translational kinetic and rotational kinetic energies. We then derive the balance of total energy and use the decomposition of energy to finally derive the balance of internal energy. 

\subsubsection{Decomposition of Energy}
We begin the decomposition of energy by rewriting the total kinetic energy arising from the particles as
\begin{equation}\label{eq:kinetic_energy_1}
\begin{split}
& \sum_{i, \alpha}  \frac{\bold{p}^{\alpha}_i \cdot \bold{p}^{\alpha}_i}{2 m^{\alpha}_i}  \Delta(\bold{x}-\bold{x}^{\alpha}_i) \\
& = \sum_{i, \alpha} \frac{1}{2} m^{\alpha}_i \Big(\frac{\bold{p}^{\alpha}_i}{m^{\alpha}_i} - \bold{v} \Big) \cdot \Big(\frac{\bold{p}^{\alpha}_i}{m^{\alpha}_i} -\bold{v} \Big) \Delta(\bold{x}-\bold{x}^{\alpha}_i) +\frac{\rho \bold{v} \cdot \bold{v}}{2} \\
&  =\sum_{i, \alpha}\frac{1}{2}  m^{\alpha}_i  \left[ \frac{\bold{p}^{\alpha}_i}{m^{\alpha}_i} -\hat{\bold{v}}^{\alpha}_i + \boldsymbol{\omega}\times (\bold{x}^{\alpha}_i-\bold{x}) \right] \cdot \left[\frac{\bold{p}^{\alpha}_i}{m^{\alpha}_i}-\hat{\bold{v}}^{\alpha}_i+ \boldsymbol{\omega}\times (\bold{x}^{\alpha}_i-\bold{x}) \right] \Delta(\bold{x}-\bold{x}^{\alpha}_i ) +\frac{\rho \bold{v} \cdot \bold{v}}{2} \\
& = \frac{1}{2}\sum_{i, \alpha} m^{\alpha}_i \Big(  \frac{\bold{p}^{\alpha}_i}{m^{\alpha}_i} -\hat{\bold{v}}^{\alpha}_i \Big) \cdot \Big(  \frac{\bold{p}^{\alpha}_i}{m^{\alpha}_i} -\hat{\bold{v}}^{\alpha}_i \Big)  \Delta(\bold{x}-\bold{x}^{\alpha}_i ) + \sum_{i, \alpha}m^{\alpha}_i \left[ \boldsymbol{\omega} \times (\bold{x}^{\alpha}_i-\bold{x}) \right] \cdot \Big(  \frac{\bold{p}^{\alpha}_i}{m^{\alpha}_i}-\bold{v}  \Big) \Delta(\bold{x}-\bold{x}^{\alpha}_i ) \\
& \hspace{125pt} -\frac{1}{2}\sum_{i, \alpha}m^{\alpha}_i\left[ \boldsymbol{\omega} \times (\bold{x}^{\alpha}_i-\bold{x})  \right] \cdot \left[ \boldsymbol{\omega} \times (\bold{x}^{\alpha}_i-\bold{x})  \right] \Delta(\bold{x}-\bold{x}^{\alpha}_i ) + \frac{\rho \bold{v} \cdot \bold{v}}{2}  \ ,
\end{split}
\end{equation}
where
\begin{equation}\label{eq:v_hat}
\hat{\bold{v}}^{\alpha}_i=\bold{v}+\boldsymbol{\omega} \times (\bold{x}^{\alpha}_i-\bold{x}) \ ,
\end{equation}
is the rigid-body  convective velocity of a particle that is translating and rotating with the continuum point $\bold{x}$. The third term on the right hand side of the third equality in \eqref{eq:kinetic_energy_1} can be rewritten as
\begin{equation}\label{eq:en_simp_1}
\begin{split}
\frac{1}{2}\sum_{i, \alpha}m^{\alpha}_i\left[ \boldsymbol{\omega} \times (\bold{x}^{\alpha}_i-\bold{x})  \right] \cdot \left[ \boldsymbol{\omega} \times (\bold{x}^{\alpha}_i-\bold{x})  \right] \Delta(\bold{x}-\bold{x}^{\alpha}_i ) &=\frac{1}{2}\bold{I}: \boldsymbol{\omega} \otimes \boldsymbol{\omega} =\frac{1}{2}\bold{I}\boldsymbol{\omega} \cdot \boldsymbol{\omega} \ ,
\end{split}
\end{equation}
where $``:"$ denotes the double contraction between two second order tensors, which for any two general tensors $\Sb$ and $\Hb$ yields a scalar $S_{ij}H_{ij}$ with  Einstein's summation convention. In obtaining \eqref{eq:en_simp_1}, we have used the identity
\begin{equation}\label{eq:en_simp_2}
\left[ \boldsymbol{\omega}  \times (\bold{x}^{\alpha}_i-\bold{x} ) \right]  \cdot \left[ \boldsymbol{\omega}  \times (\bold{x}^{\alpha}_i-\bold{x}) \right]  = \Big\{ \Big[(\bold{x}^{\alpha}_i-\bold{x}) \cdot (\bold{x}^{\alpha}_i-\bold{x}) \Big]\bold{i} - \Big[(\bold{x}^{\alpha}_i-\bold{x}) \otimes (\bold{x}^{\alpha}_i-\bold{x})\Big] \Big\} : \boldsymbol{\omega}  \otimes \boldsymbol{\omega}  \ .
\end{equation}
The second term on the right hand side of the third equality in \eqref{eq:kinetic_energy_1} can be rewritten as
\begin{equation}\label{eq:en_simp_3}
\begin{split}
\sum_{i, \alpha}m^{\alpha}_i \left[ \boldsymbol{\omega} \times (\bold{x}^{\alpha}_i-\bold{x}) \right] \cdot \Big(  \frac{\bold{p}^{\alpha}_i}{m^{\alpha}_i}-\bold{v} \Big) \Delta(\bold{x}-\bold{x}^{\alpha}_i ) &=\sum_{i, \alpha} m^{\alpha}_i \Big[ (\bold{x}^{\alpha}_i-\bold{x}) \times \left(\frac{\bold{p}^{\alpha}_i}{m^{\alpha}_i}-\bold{v}\right) \Big] \cdot \boldsymbol{\omega} \Delta(\bold{x}-\bold{x}^{\alpha}_i ) \\
&=\rho \bold{M} \cdot \boldsymbol{\omega} \\
&=\bold{I}\boldsymbol{\omega} \cdot \boldsymbol{\omega} \ ,
\end{split}
\end{equation}
using the definition of spin angular momentum \eqref{eq:def_omega} and an assumption that the continuum point $\xb$ represents the center of mass of particles in its neighborhood defined by the length scale of averaging in the coarse-graining function, \emph{i.e.}, 
\begin{equation}\label{eq:com}
\rho \xb = \sum_{i,\alpha} m^\alpha_i \xb^\alpha_i \Delta(\xb - \xb^\alpha_i) \ . 
\end{equation}
Making use of \eqref{eq:en_simp_1} and \eqref{eq:en_simp_3}, the kinetic energy, in equation \eqref{eq:kinetic_energy_1}, can be reduced to
\begin{equation}\label{eq:en_simp_4}
\sum_{i, \alpha} \frac{\bold{p}^{\alpha}_i \cdot \bold{p}^{\alpha}_i}{2 m^{\alpha}_i} \Delta(\bold{x}-\bold{x}^{\alpha}_i) = \frac{1}{2}\sum_{i, \alpha} m^{\alpha}_i \Big(  \frac{\bold{p}^{\alpha}_i}{m^{\alpha}_i} -\hat{\bold{v}}^{\alpha}_i \Big) \cdot \Big(  \frac{\bold{p}^{\alpha}_i}{m^{\alpha}_i} -\hat{\bold{v}}^{\alpha}_i \Big)  \Delta(\bold{x}-\bold{x}^{\alpha}_i ) + \frac{\rho \bold{v} \cdot \bold{v}}{2}  + \frac{\bold{I \boldsymbol{\omega}} \cdot \boldsymbol{\omega}}{2} \ .
\end{equation}
Using \eqref{eq:en_simp_4}, the total energy at the macroscale \eqref{eq:total_energy} can be simplified to 
\begin{equation}
\begin{split}
\label{eq:en_simp_5}
\rho e &= \frac{1}{2}\sum_{i, \alpha, j, \beta} u_2(\bold{x}^{\alpha}_i,\bold{x}^{\beta}_j) \Delta(\bold{x}-\bold{x}^{\alpha}_i)+\frac{1}{2}\sum_{i, \alpha} u^{s}(\bold{x}^1_i,\bold{x}^2_i)\Delta(\bold{x}-\bold{x}^{\alpha}_i) \\
&  \hspace{28pt}+\frac{1}{2}\sum_{i, \alpha}m^{\alpha}_i(\bold{v}^{\alpha}_i-\hat{\bold{v}}^{\alpha}_i) \cdot (\bold{v}^{\alpha}_i-\hat{\bold{v}}^{\alpha}_i) \Delta (\bold{x}-\bold{x}^{\alpha}_i)  + \frac{\rho \bold{v} \cdot \bold{v}}{2}+\frac{\bold{I}\boldsymbol{\omega}\cdot \boldsymbol{\omega}}{2} \ .
\end{split}
\end{equation}
Denoting the first three terms on the right hand side of \eqref{eq:en_simp_5} as the total internal energy $\rho(\xb,t) \epsilon(\xb,t)$ at the macroscopic point $\xb$, given by 
\begin{equation}
\begin{split}
\label{eq:total_energy_internal}
\rho \epsilon &= \frac{1}{2}\sum_{i, \alpha, j, \beta} u_2(\bold{x}^{\alpha}_i,\bold{x}^{\beta}_j) \Delta(\bold{x}-\bold{x}^{\alpha}_i)+\frac{1}{2}\sum_{i, \alpha} u^{s}(\bold{x}^1_i,\bold{x}^2_i)\Delta(\bold{x}-\bold{x}^{\alpha}_i) \\
&  \hspace{28pt}+\frac{1}{2}\sum_{i, \alpha}m^{\alpha}_i(\bold{v}^{\alpha}_i-\hat{\bold{v}}^{\alpha}_i) \cdot (\bold{v}^{\alpha}_i-\hat{\bold{v}}^{\alpha}_i) \Delta (\bold{x}-\bold{x}^{\alpha}_i) \ ,
\end{split}
\end{equation}
the total energy \eqref{eq:total_energy} can be decomposed into internal energy and translational and rotational kinetic energies as 
\begin{equation}
\label{eq:total_energy_short}
\rho e = \rho\epsilon + \frac{\rho \bold{v} \cdot \bold{v}}{2}+\frac{\bold{I}\boldsymbol{\omega}\cdot \boldsymbol{\omega}}{2} \ .
\end{equation}

\subsubsection{Balance of Total Energy}

Taking the time derivative of both sides of \eqref{eq:total_energy}, we have
\begin{equation}
 \begin{split}
 \label{eq:energy_deriv}
\dot{\rho} e + \rho \dot{ e}  &= \sum_{i, \alpha} \frac{\bold{p}^{\alpha}_i \cdot \dot{\bold{p}}^{\alpha}_i}{m^{\alpha}_i}\Delta(\bold{x}-\bold{x}^{\alpha}_i) + \sum_{i ,\alpha} E^{\alpha}_i \left[ \frac{\partial \Delta (\bold{x}-\bold{x}^{\alpha}_i)}{\partial \bold{x}} \cdot \left( \bold{v} - \frac{\bold{p}^{\alpha}_i}{m^{\alpha}_i}  \right) \right] \\
&  \hspace{28pt} +\frac{1}{2}\sum_{i, \alpha, j, \beta} \left[ \frac{\partial u_2 (\bold{x}^{\alpha}_i,\bold{x}^{\beta}_j)}{\partial \bold{x}^{\alpha}_i} \cdot \frac{\bold{p}^{\alpha}_i}{m^{\alpha}_i}+ \frac{\partial u_2 (\bold{x}^{\alpha}_i,\bold{x}^{\beta}_j)}{\partial \bold{x}^{\beta}_j} \cdot \frac{\bold{p}^{\beta}_j}{m^{\beta}_j} \right] \Delta(\bold{x}-\bold{x}^{\alpha}_i) \\
&  \hspace{56pt} +\frac{1}{2}\sum_{i, \alpha} \left[ \frac{\partial u_s(\bold{x}^1_i,\bold{x}^2_i)}{\partial \bold{x}^1_i} \cdot \frac{\bold{p}^{1}_i}{m^{1}_i}+\frac{\partial u_s(\bold{x}^1_i,\bold{x}^2_i)}{\partial \bold{x}^2_i} \cdot \frac{\bold{p}^2_i}{m^2_i} \right] \Delta (\bold{x}-\bold{x}^{\alpha}_i) \ ,
\end{split}
\end{equation}
where $E^{\alpha}_i$ is the total energy of each atom given by 
\begin{equation}\label{eq:atom_en}
E^{\alpha}_i = \frac{\bold{p}^{\alpha}_i \cdot \bold{p}^{\alpha}_i}{2 m^{\alpha}_i} + \frac{1}{2}\sum_{j, \beta} u_2(\bold{x}^{\alpha}_i,\bold{x}^{\beta}_j)+\frac{1}{2}u_s (\bold{x}^1_i,\bold{x}^2_i) \ .
\end{equation}
Note that the interaction energy from the pair potential and the harmonic spring energy is equally divided between two interacting particles. 
Using equations \eqref{eq:mass_bal} and \eqref{eq:motion}, we can rewrite \eqref{eq:energy_deriv} as
\begin{equation}
\begin{split}
\label{eq:energy_deriv2}
\rho\dot{e}-\rho \left( \frac{\partial}{\partial \bold{x}} \cdot \bold{v} \right) e &= \sum_{i, \alpha} \frac{\bold{p}^{\alpha}_i}{m^{\alpha}_i} \cdot \left[ - \zeta \frac{\bold{p}^{\alpha} _i}{m_i} +  \sum_{j, \beta} \bold{F}^{\alpha \beta}_{ij} +\bold{f}^{\alpha}_i-\frac{\partial u_s(\bold{x}^{1}_i, \bold{x}^{2}_i )}{\partial \bold{x}^{\alpha}_i} +\sqrt{2\kt \zeta}\, \frac{\mathrm{d}\bold{W}}{\mathrm{d}t} \right]  \Delta (\bold{x}-\bold{x}^{\alpha}_i) \\
& \hspace{28pt}+ \sum_{i ,\alpha} E^{\alpha}_i \left[ \frac{\partial \Delta (\bold{x}-\bold{x}^{\alpha}_i)}{\partial \bold{x}} \cdot \left( \bold{v} - \frac{\bold{p}^{\alpha}_i}{m^{\alpha}_i}  \right) \right] \\
&  \hspace{56pt}+\frac{1}{2}\sum_{i, \alpha, j, \beta} \left[ \frac{\partial u_2 (\bold{x}^{\alpha}_i,\bold{x}^{\beta}_j)}{\partial \bold{x}^{\alpha}_i} \cdot \frac{\bold{p}^{\alpha}_i}{m^{\alpha}_i}+ \frac{\partial u_2 (\bold{x}^{\alpha}_i,\bold{x}^{\beta}_j)}{\partial \bold{x}^{\beta}_j} \cdot \frac{\bold{p}^{\beta}_j}{m^{\beta}_j} \right] \Delta(\bold{x}-\bold{x}^{\alpha}_i) \\
&  \hspace{84pt} +\frac{1}{2}\sum_{i, \alpha} \left[ \frac{\partial u_s(\bold{x}^1_i,\bold{x}^2_i)}{\partial \bold{x}^1_i} \cdot \frac{\bold{p}^{1}_i}{m^{1}_i}+\frac{\partial u_s(\bold{x}^1_i,\bold{x}^2_i)}{\partial \bold{x}^2_i} \cdot \frac{\bold{p}^2_i}{m^2_i} \right] \Delta (\bold{x}-\bold{x}^{\alpha}_i) \ .
\end{split}
\end{equation}

We now simplify each term in \eqref{eq:energy_deriv2}. To this end, the second term on the right hand side of \eqref{eq:energy_deriv2} can be rewritten as
\begin{equation}\label{eq:tot_sim_1}
\begin{split}
\sum_{i, \alpha} E^{\alpha}_i \left[ \frac{\partial \Delta (\bold{x}-\bold{x}^{\alpha}_i)}{\partial \bold{x}} \cdot \left( \bold{v} - \frac{\bold{p}^{\alpha}_i}{m^{\alpha}_i}  \right) \right]
&= \frac{\partial}{\partial \bold{x}} \left( \sum_{i, \alpha} E^{\alpha}_i \Delta (\bold{x}-\bold{x}^{\alpha}_i) \right) \cdot \bold{v} -\frac{\partial}{\partial \bold{x}} \cdot \left( \sum_{i,\alpha} E^{\alpha}_i \frac{\bold{p}^{\alpha}_i}{m^{\alpha}_i} \Delta (\bold{x}-\bold{x}^{\alpha}_i) \right) \\
 &=-\rho \left( \frac{\partial}{\partial \bold{x}} \cdot \bold{v}\right) e -  \frac{\partial}{\partial \bold{x}} \cdot \left( \sum_{i,\alpha} E^{\alpha}_i \left(\frac{\bold{p}^{\alpha}_i}{m^{\alpha}_i} - \vb \right) \Delta (\bold{x}-\bold{x}^{\alpha}_i) \right) \ ,
\end{split}
\end{equation}
where use is made of the relation \eqref{eq:total_energy}. Equation \eqref{eq:tot_sim_1} can be further simplified by manipulating the kinetic energy part of the atomic energy $E_i^\alpha$ in \eqref{eq:atom_en} to yield
%
\begin{align}
\sum_{i, \alpha} & E^{\alpha}_i \left[ \frac{\partial \Delta (\bold{x}-\bold{x}^{\alpha}_i)}{\partial \bold{x}} \cdot \left( \bold{v} - \frac{\bold{p}^{\alpha}_i}{m^{\alpha}_i}  \right) \right] \nonumber \\
 & =-\rho \left( \frac{\partial}{\partial \bold{x}} \cdot \bold{v}\right) e - \frac{\partial}{\partial \bold{x}} \cdot \Bigg(\sum_{i, \alpha} \bigg[ \frac{(\bold{p}^{\alpha}_i-m^{\alpha}_i \bold{v})\cdot (\bold{p}^{\alpha}_i-m^{\alpha}_i \bold{v})}{2m^{\alpha}_i} +   \frac{1}{2}\sum_{j, \beta} u_2(\bold{x}^{\alpha}_i,\bold{x}^{\beta}_j) \nonumber \\ 
& \hspace{1.5in}+\frac{1}{2}u_s (\bold{x}^1_i,\bold{x}^2_i) \bigg] \Big( \frac{\bold{p}^{\alpha}_i}{m^{\alpha}_i} -\bold{v} \Big) \Delta (\bold{x}-\bold{x}^{\alpha}_i) \Bigg) + \frac{\partial}{\partial \bold{x}} \cdot \left( \Big(\bold{T}^{\text{K}} \Big)^T \bold{v} \right)\ ,
\end{align}
%
where the last term represents the contribution from the kinetic part of the stress tensor \eqref{eq:kinetic_stress}.

Next, the third term on the right hand side of \eqref{eq:energy_deriv2} can be rewritten as 
\begin{equation}\label{eq:tot_sim_2}
\begin{split}
&\frac{1}{2}\sum_{i, \alpha, j, \beta} \left[ \frac{\partial u_2 (\bold{x}^{\alpha}_i,\bold{x}^{\beta}_j)}{\partial \bold{x}^{\alpha}_i} \cdot \frac{\bold{p}^{\alpha}_i}{m^{\alpha}_i}+ \frac{\partial u_2 (\bold{x}^{\alpha}_i,\bold{x}^{\beta}_j)}{\partial \bold{x}^{\beta}_j} \cdot \frac{\bold{p}^{\beta}_j}{m^{\beta}_j} \right] \Delta(\bold{x}-\bold{x}^{\alpha}_i) \\
&\hspace{0.5in} = - \frac{1}{2}\sum_{i, \alpha, j, \beta} \left( \bold{F}^{\alpha \beta}_{ij} \cdot \frac{\bold{p}^{\alpha}_i}{m^{\alpha}_i} \right) \Delta (\bold{x}-\bold{x}^{\alpha}_i) \\
& \hspace{1.25in} + \frac{1}{2}\sum_{i, \alpha, j, \beta}  \left( \frac{\partial  u_2 (\bold{x}^{\alpha}_i,\bold{x}^{\beta}_i)}{\partial \bold{x}^{\alpha}_i} \cdot \frac{\bold{p}^{\alpha}_i}{m^{\alpha}_i}\right) \left(\Delta(\bold{x}-\bold{x}^{\beta}_j)+\Delta(\bold{x}-\bold{x}^{\alpha}_i)-\Delta(\bold{x}-\bold{x}^{\alpha}_i) \right)  \\
&\hspace{0.5in}=-\sum_{i, \alpha, j, \beta} \left( \bold{F}^{\alpha \beta}_{ij} \cdot \frac{\bold{p}^{\alpha}_i}{m^{\alpha}_i} \right) \Delta(\bold{x}-\bold{x}^{\alpha}_i)-\frac{1}{2}\sum_{i, \alpha, j, \beta} \left( \bold{F}^{\alpha \beta}_{ij} \cdot \frac{\bold{p}^{\alpha}_i}{m^{\alpha}_i} \right) \left( \frac{\partial}{\partial \bold{x}} \cdot \left(\bold{x}^{\alpha \beta}_{ij} b^{\alpha \beta}_{ij} \right) \right) \\
& \hspace{0.5in} = - \sum_{i, \alpha, j, \beta} \left( \bold{F}^{\alpha \beta}_{ij} \cdot \frac{\bold{p}^{\alpha}_i}{m^{\alpha}_i} \right) \Delta (\bold{x}-\bold{x}^{\alpha}_i)  - \frac{\partial}{\partial \bold{x}} \cdot \left( \frac{1}{2}\sum_{i,\alpha,j,\beta} \left[ \bold{F}^{\alpha \beta}_{ij} \cdot \Big(\frac{\bold{p}^{\alpha} _i}{m_i} - \bold{v}  \Big) \right] \bold{x}^{\alpha \beta}_{ij} b^{\alpha \beta}_{ij}  \right) \\ 
 & \hspace{2.75in}+ \frac{\partial}{\partial \bold{x}} \cdot \left( \left(\bold{T}^{\text{V}} \right)^T \bold{v} \right) \ ,
\end{split}
\end{equation}
where $\bold{T}^{\text{V}}$ is the virial stress in \eqref{eq:virial_stress}.

Employing similar procedures in deriving \eqref{eq:tot_sim_2}, the fourth term on the right hand side of \eqref{eq:energy_deriv2} can be modified to yield 
\begin{equation}\label{eq:tot_sim_3}
\begin{split}
&\frac{1}{2}\sum_{i, \alpha} \left( \frac{\partial u_s(\bold{x}^1_i,\bold{x}^2_i)}{\partial \bold{x}^1_i} \cdot \frac{\bold{p}^{1}_i}{m^{1}_i}+\frac{\partial u_s(\bold{x}^1_i,\bold{x}^2_i)}{\partial \bold{x}^2_i} \cdot \frac{\bold{p}^2_i}{m^2_i} \right) \Delta (\bold{x}-\bold{x}^{\alpha}_i) \\
& \hspace{0.05in} = \sum_i \left[ \left( \frac{\partial u_s(\bold{x}^1_i,\bold{x}^2_i)}{\partial \bold{x}^1_i} \cdot \frac{\bold{p}^{1}_i}{m^{1}_i} \right) \Delta(\bold{x}-\bold{x}^{1}_i) +  \left(\frac{\partial u_s(\bold{x}^1_i,\bold{x}^2_i)}{\partial \bold{x}^2_i} \cdot \frac{\bold{p}^2_i}{m^2_i} \right) \Delta(\bold{x}-\bold{x}^{2}_i)  \right] \\
& \hspace{0.05in}  =  \frac{\partial}{\partial \bold{x}} \cdot \left( \frac{1}{2} \sum_i \left( \left[ \frac{\partial u_s(\bold{x}^1_i,\bold{x}^2_i)}{\partial \bold{x}^1_i} \cdot \left( \frac{\bold{p}^{1} _i}{m_i} - \bold{v} \right) \right] \bold{x}^{12}_{ii}b^{12}_{ii} + \left[ \frac{\partial u_s(\bold{x}^1_i,\bold{x}^2_i)}{\partial \bold{x}^2_i} \cdot \left( \frac{\bold{p}^{2} _i}{m_i} - \bold{v}  \right) \right]\bold{x}^{21}_{ii}b^{21}_{ii} \right)  \right)  \\
&\hspace{01.5in}+ \frac{\partial}{\partial \bold{x}} \cdot \left( \left(\bold{T}^{\text{S}}\right)^T \bold{v} \right) \ ,
\end{split}
\end{equation}
where $\bold{T}^{\text{S}}$ is the stress due to the harmonic spring terms given by \eqref{eq:spring_stress}. 

The active forces term in \eqref{eq:energy_deriv2} can be simplified as 
\begin{equation}\label{eq:tot_sim_4}
\begin{split}
\sum_{i,\alpha} \frac{\bold{p}^{\alpha}_i}{m^{\alpha}_i} \cdot \bold{f}^{\alpha}_i \Delta (\bold{x}-\bold{x}^{\alpha}_i) 
&= \sum_{i} \left( \frac{\bold{p}^1_i}{m^1_i} -  \frac{\bold{p}^2_i}{m^2_i}  \right) \cdot \bold{f}_i \Delta(\bold{x}-\bold{x}^{1}_i) \\ 
& \hspace{0.25in} - \frac{\partial}{\partial \bold{x}} \cdot \left( \sum_i  \left[ \bold{f}_i \cdot \left( \frac{\bold{p}^2_i}{m^2_i} -\bold{v} \right) \right] \bold{x}^{12}_{ii}b^{12}_{ii} \right) + \frac{\partial}{\partial \bold{x}} \cdot \left( \left(\bold{T}^{\text{A}} \right)^T \bold{v}  \right) \ ,
\end{split}
\end{equation}
where $\bold{T}^{\text{A}}$ is the active stress in \eqref{eq:active_stress}. 

Combining the results from \eqref{eq:tot_sim_1} to \eqref{eq:tot_sim_4} and rearranging the terms, the total energy balance in \eqref{eq:energy_deriv2} reduces to
\begin{equation}
\begin{split}
\label{eq:third_energy_deriv}
\rho\dot{e}&=\sum_{i,\alpha} \frac{\bold{p}^{\alpha}_i}{m^{\alpha}_i} \cdot \left( - \zeta \frac{\bold{p}^{\alpha} _i}{m_i} +  \sqrt{2\kt \zeta}\, \frac{\mathrm{d}\bold{W}}{\mathrm{d}t}  \right) \Delta (\bold{x}-\bold{x}^{\alpha}_i)+ \frac{\partial}{\partial \bold{x}} \cdot \Big( \bold{T}^T \bold{v} \Big) + \sum_i \Big( \frac{\bold{p}^1_i}{m^1_i}-\frac{\bold{p}^2_i}{m^2_i} \Big) \cdot \bold{f}_i \Delta (\bold{x}-\bold{x}^{\alpha}_i)  \\
& - \frac{\partial}{\partial \bold{x}} \cdot \Bigg(\sum_{i, \alpha} \bigg[ \frac{(\bold{p}^{\alpha}_i-m^{\alpha}_i \bold{v})\cdot (\bold{p}^{\alpha}_i-m^{\alpha}_i \bold{v})}{2m^{\alpha}_i} +   \frac{1}{2}\sum_{j, \beta} u_2(\bold{x}^{\alpha}_i,\bold{x}^{\beta}_j) +\frac{1}{2}u_s (\bold{x}^1_i,\bold{x}^2_i) \bigg] \Big( \frac{\bold{p}^{\alpha}_i}{m^{\alpha}_i} -\bold{v} \Big) \Delta (\bold{x}-\bold{x}^{\alpha}_i) \Bigg) \\
& \hspace{0.15in} - \frac{\partial}{\partial \bold{x}} \cdot \left( \frac{1}{2}\sum_{i,\alpha,j,\beta} \left[ \bold{F}^{\alpha \beta}_{ij} \cdot \Big(\frac{\bold{p}^{\alpha} _i}{m_i} - \bold{v}  \Big) \right] \bold{x}^{\alpha \beta}_{ij} b^{\alpha \beta}_{ij}  \right) 
- \frac{\partial}{\partial \bold{x}} \cdot \left( \sum_i  \left[ \bold{f}_i \cdot \left( \frac{\bold{p}^2_i}{m^2_i} -\bold{v} \right) \right] \bold{x}^{12}_{ii}b^{12}_{ii} \right)  \\ 
&  \hspace{0.25in} + \frac{\partial}{\partial \bold{x}} \cdot \left( \frac{1}{2} \sum_i \left( \left[ \frac{\partial u_s(\bold{x}^1_i,\bold{x}^2_i)}{\partial \bold{x}^1_i} \cdot \left( \frac{\bold{p}^{1} _i}{m_i} - \bold{v} \right) \right] \bold{x}^{12}_{ii}b^{12}_{ii} + \left[ \frac{\partial u_s(\bold{x}^1_i,\bold{x}^2_i)}{\partial \bold{x}^2_i} \cdot \left( \frac{\bold{p}^{2} _i}{m_i} - \bold{v}  \right) \right]\bold{x}^{21}_{ii}b^{21}_{ii} \right)  \right) \ , 
\end{split}
\end{equation}
where the last four terms contain divergence terms that are the fluxes of energy in the system. 

At this stage, it can be seen that the second term on the right hand side of the energy balance in \eqref{eq:third_energy_deriv} contains the rate of work done due to the applied forces in terms of the stress tensor. What remains to be seen is the form for the rate of work performed by the surface couples in the system. To this end, we modify each of the last four terms of \eqref{eq:third_energy_deriv} by subtracting the rotational parts of the velocity from the individual atomic velocities. 

Beginning with the fifth term on the right hand side of \eqref{eq:third_energy_deriv}, it can be seen that 
\begin{equation}\label{eq:heat_1}
\begin{split}
 \sum_{i, \alpha, j, \beta} \frac{1}{2} & \left[ \bold{F}^{\alpha \beta}_{ij}  \cdot \left( \frac{\bold{p}^{\alpha}_i}{m^{\alpha}_i} -\bold{v} \right) \right]  \bold{x}^{\alpha \beta}_{ij} b^{\alpha \beta}_{ij} \\
 & = \sum_{i, \alpha, j, \beta} \frac{1}{2} \left[ \bold{F}^{\alpha \beta}_{ij} \cdot \left( \frac{\bold{p}^{\alpha}_i}{m^{\alpha}_i} -\hat{\bold{v}}_i^\alpha \right) \right] \bold{x}^{\alpha \beta}_{ij} b^{\alpha \beta}_{ij} 
 +\sum_{i, \alpha, j, \beta} \frac{1}{2} \left( \bold{F}^{\alpha \beta}_{ij} \cdot \left[ \boldsymbol{\omega} \times \left( \bold{x}^{\alpha}_i - \bold{x} \right) \right] \right) \bold{x}^{\alpha \beta}_{ij} b^{\alpha \beta}_{ij} \\
 & = \sum_{i, \alpha, j, \beta} \frac{1}{2} \left[ \bold{F}^{\alpha \beta}_{ij} \cdot \left( \frac{\bold{p}^{\alpha}_i}{m^{\alpha}_i} -\hat{\bold{v}}_i^\alpha \right) \right] \bold{x}^{\alpha \beta}_{ij} b^{\alpha \beta}_{ij}  
  +\sum_{i, \alpha, j, \beta} \left[ \frac{1}{2} \bold{x}^{\alpha \beta}_{ij} \otimes \left( \left( \bold{x}^{\alpha}_i - \bold{x} \right) \times \bold{F}^{\alpha \beta}_{ij}  b^{\alpha \beta}_{ij} \right)\right]  \boldsymbol{\omega} \\
 &=\bold{J}^{\text{V}}_\text{q}-(\bold{C}^{\text{V}})^{T} \boldsymbol{\omega} \ ,
 \end{split} 
 \end{equation}
where 
\begin{equation}\label{eq:virial_heat}
\bold{J}^\text{V}_{\text{q}} =  \sum_{i, \alpha, j, \beta} \frac{1}{2} \left[ \bold{F}^{\alpha \beta}_{ij} \cdot \left( \frac{\bold{p}^{\alpha}_i}{m^{\alpha}_i} -\hat{\bold{v}}_i^\alpha \right) \right] \bold{x}^{\alpha \beta}_{ij} b^{\alpha \beta}_{ij}  \ ,
\end{equation}
and $\bold{C}^{\text{V}}$ is the virial part of the couple stress given by \eqref{eq:couple_stress_V}. Following similar procedures used to obtain the equations \eqref{eq:heat_1} and \eqref{eq:virial_heat}, the last term on the right hand side of \eqref{eq:third_energy_deriv} is reduced to 
\begin{equation}\label{eq:heat_2}
\begin{split}
& \frac{1}{2} \sum_i \left( \left[ \frac{\partial u_s(\bold{x}^1_i,\bold{x}^2_i)}{\partial \bold{x}^1_i} \cdot \left( \frac{\bold{p}^{1} _i}{m_i} - \bold{v} \right) \right] \bold{x}^{12}_{ii}b^{12}_{ii} + \left[ \frac{\partial u_s(\bold{x}^1_i,\bold{x}^2_i)}{\partial \bold{x}^2_i} \cdot \left( \frac{\bold{p}^{2} _i}{m_i} - \bold{v}  \right) \right]\bold{x}^{21}_{ii}b^{21}_{ii} \right)   \\
 & \hspace{3in} =- \bold{J}^{\text{S}}_\text{q}-(\bold{C}^{\text{S}})^{T} \boldsymbol{\omega} \ ,
 \end{split} 
 \end{equation}
where 
\begin{equation}\label{eq:spring_heat}
\bold{J}^\text{S}_{\text{q}} =  - \sum_i \left[ \frac{\partial u_s(\bold{x}^1_i,\bold{x}^2_i)}{\partial \bold{x}^1_i}  \cdot \left( \frac{\bold{p}^1_i}{m^1_i} -\hat{\bold{v}}_i^1 \right) \bold{x}^{12}_{ii} b^{12}_{ii}+\frac{\partial u_s(\bold{x}^1_i,\bold{x}^2_i)}{\partial \bold{x}^2_i} \cdot \left( \frac{\bold{p}^2_i}{m^2_i} -\hat{\bold{v}}_i^2 \right) \bold{x}^{21}_{ii} b^{21}_{ii} \right] \ ,
\end{equation}
and $\bold{C}^{\text{S}}$ is the spring part of the couple stress given by \eqref{eq:couple_stress_S}.

Next, the active terms contained in the sixth term on the right hand side of \eqref{eq:third_energy_deriv} can be reduced to 
\begin{equation}\label{eq:heat_3}
\begin{split}
\sum_i  \left[ \bold{f}_i \cdot \left( \frac{\bold{p}^2_i}{m^2_i} -\bold{v} \right) \right] \bold{x}^{12}_{ii}b^{12}_{ii}   \hspace{0in} & =
\sum_i \left[ \bold{f}_i \cdot \left( \frac{\bold{p}^2_i}{m^2_i} -\hat{\bold{v}}_i^2 \right) \right] \bold{x}^{12}_{ii}b^{12}_{ii} + \sum_i \left[ \fb_i \cdot \left( \boldsymbol{\omega} \times (\bold{x}^2_i-\bold{x}) \right)  \right] \bold{x}^{12}_{ii} b^{12}_{ii} \\
& = \bold{J}^{\text{A}}_\text{q}-(\bold{C}^{\text{A}})^{T} \boldsymbol{\omega} \ ,
 \end{split}
\end{equation} 
where
\begin{equation}\label{eq:active_heat}
\bold{J}^{\text{A}}_\text{q} = \sum_i \left[ \bold{f}_i \cdot \left( \frac{\bold{p}^2_i}{m^2_i} -\hat{\bold{v}}_i^2 \right) \right] \bold{x}^{12}_{ii}b^{12}_{ii} \ , 
\end{equation}
and $\bold{C}^{\text{A}}$ is the active part of the couple stress given by \eqref{eq:couple_stress_A}.

The fourth term on the right-hand side of \eqref{eq:third_energy_deriv} can be manipulated by subtracting and adding the  rigid rotational components of the velocity $\boldsymbol{\omega} \times (\bold{x}^{\alpha}_i-\bold{x})$  from the relative kinetic energy term to yield 
\begin{equation}\label{eq:heat_4}
\begin{split}
& \sum_{i, \alpha} \bigg[ \frac{(\bold{p}^{\alpha}_i-m^{\alpha}_i \bold{v})\cdot (\bold{p}^{\alpha}_i-m^{\alpha}_i \bold{v})}{2m^{\alpha}_i} +   \frac{1}{2}\sum_{j, \beta} u_2(\bold{x}^{\alpha}_i,\bold{x}^{\beta}_j) +\frac{1}{2}u_s (\bold{x}^1_i,\bold{x}^2_i) \bigg] \Big( \frac{\bold{p}^{\alpha}_i}{m^{\alpha}_i} -\bold{v} \Big) \Delta (\bold{x}-\bold{x}^{\alpha}_i) \\ 
& = \sum_{i, \alpha} \left[ \hat{\bold{K}}^{\alpha}_i +\frac{1}{2}\sum_{j, \beta} u_2(\bold{x}^{\alpha}_i,\bold{x}^{\beta}_j) +\frac{1}{2}u_s (\bold{x}^1_i,\bold{x}^2_i) - m_i^\alpha \vb \cdot \left[\boldsymbol{\omega} \times (\bold{x}^{\alpha}_i-\bold{x})\right]  \right] \left( \frac{\bold{p}^{\alpha}_i}{m^{\alpha}_i} -  \bold{v} \right) \Delta (\bold{x}-\bold{x}^{\alpha}_i)  \\
&  \hspace{1.5 in}  + \sum_{i, \alpha} m^{\alpha}_i \left( \left[\boldsymbol{\omega} \times (\bold{x}^{\alpha}_i - \bold{x})\right] \cdot  \frac{\bold{p}^{\alpha}_i}{m^{\alpha}_i}  \right) \left( \frac{\bold{p}^{\alpha}_i}{m^{\alpha}_i} - \bold{v} \right)  \Delta (\bold{x}-\bold{x}^{\alpha}_i) \\
 & \hspace{2in} -\frac{1}{2} \sum_{i, \alpha} \left( \bold{I}^{\alpha}_i \boldsymbol{\omega} \cdot \boldsymbol{\omega} \right) \left( \frac{\bold{p}^{\alpha}_i}{m^{\alpha}_i} - \bold{v} \right)  \Delta (\bold{x}-\bold{x}^{\alpha}_i) \\
&=\bold{J}^{\text{K}}_\text{q}-(\bold{C}^{\text{K}})^{T} \boldsymbol{\omega}  + \frac{1}{2}\left(\bold{Y}:  \boldsymbol{\omega} \otimes \boldsymbol{\omega} \right) \ ,
 \end{split} 
 \end{equation}
where
\begin{equation}\label{eq:rel_kin_rot_tra}
\hat{\bold{K}}^{\alpha}_i =  \frac{1}{2}m^{\alpha}_i\left( \frac{\bold{p}^{\alpha}_i}{m^{\alpha}_i} - \hat{\bold{v}}_i^\alpha \right) \cdot \left( \frac{\bold{p}^{\alpha}_i}{m^{\alpha}_i} - \hat{\bold{v}}_i^\alpha\right)
 \end{equation}
is the kinetic energy of the particle relative to the translational and the rotational motion of the continuum point, 
 \begin{equation}\label{eq:kinetic_heat}
\bold{J}^{\text{K}}_\text{q} = \sum_{i \alpha}\left( \hat{\bold{K}}^{\alpha}_i + \frac{1}{2}\sum_{j, \beta} u_2(\bold{x}^{\alpha}_i,\bold{x}^{\beta}_j) +\frac{1}{2}u_s (\bold{x}^1_i,\bold{x}^2_i) -  m_i^\alpha\bold{v} \cdot \boldsymbol{\omega} \times (\bold{x}^{\alpha}_i - \bold{x}) \right) \left( \frac{\bold{p}^{\alpha}_i}{m^{\alpha}_i} - \bold{v} \right) \Delta (\bold{x}-\bold{x}^{\alpha}_i)  \ ,
 \end{equation}
 and $\bold{C}^{\text{K}}$ is the kinetic part of the couple stress given by \eqref{eq:couple_stress_K}.
 
 With the manipulations from \eqref{eq:heat_1} - \eqref{eq:kinetic_heat}, the balance of total energy at the macroscopic point in \eqref{eq:third_energy_deriv} is reduced to 
\begin{equation}
\begin{split}
\label{eq:fourth_energy_deriv}
\rho\dot{e}&=\sum_{i,\alpha} \frac{\bold{p}^{\alpha}_i}{m^{\alpha}_i} \cdot \left( - \zeta \frac{\bold{p}^{\alpha} _i}{m_i} +  \sqrt{2\kt \zeta}\, \frac{\mathrm{d}\bold{W}}{\mathrm{d}t}  \right) \Delta (\bold{x}-\bold{x}^{\alpha}_i)  + \sum_i \Big( \frac{\bold{p}^1_i}{m^1_i}-\frac{\bold{p}^2_i}{m^2_i} \Big) \cdot \bold{f}_i \Delta (\bold{x}-\bold{x}^{\alpha}_i)  \\
& \hspace{1in}- \frac{\partial}{\partial \bold{x}} \cdot \Big( \Jb_{\text{q}} \Big) + \frac{\partial}{\partial \bold{x}} \cdot \Big( \bold{T}^T \bold{v} \Big) + \frac{\partial}{\partial \bold{x}} \cdot \Big( \bold{C}^T \bold{\omega} \Big) - \frac{1}{2}  \frac{\partial }{\partial \xb} \cdot  \left(\bold{Y}:  \boldsymbol{\omega} \otimes \boldsymbol{\omega} \right) 
\ , 
\end{split}
\end{equation}
where 
\begin{equation}\label{eq:total_heat_flux}
\bold{J}_{\text{q}} = \bold{J}^{\text{K}}_{\text{q}} + \bold{J}^{\text{V}}_{\text{q}}+\bold{J}^{\text{S}}_{\text{q}}+\bold{J}^{\text{A}}_{\text{q}} \ .
\end{equation}
We can further reduce the first term on the right hand side of \eqref{eq:fourth_energy_deriv} by adding and subtracting the terms corresponding to the rotational velocity of the particle moving with the continuum point $\boldsymbol{\omega} \times (\bold{x}^{\alpha}_i-\bold{x})$ to 
obtain
\begin{equation}\label{eq:heat_5}
\begin{split}
& \sum_{i,\alpha} \frac{\bold{p}^{\alpha}_i}{m^{\alpha}_i} \cdot \left( - \zeta \frac{\bold{p}^{\alpha} _i}{m_i} +  \sqrt{2\kt \zeta}\, \frac{\mathrm{d}\bold{W}}{\mathrm{d}t}  \right) \Delta (\bold{x}-\bold{x}^{\alpha}_i)  + \sum_i \Big( \frac{\bold{p}^1_i}{m^1_i}-\frac{\bold{p}^2_i}{m^2_i} \Big) \cdot \bold{f}_i \Delta (\bold{x}-\bold{x}^{\alpha}_i) \\ 
& \hspace{0.31in} =\sum_{i, \alpha} \left( \frac{\bold{p}^{\alpha}_i}{m^{\alpha}_i} - \hat{\bold{v}}_i^\alpha  \right) \cdot \left( -\zeta \frac{\bold{p}^{\alpha}_i}{m^{\alpha}_i} +\sqrt{2k_BT \zeta} \frac{d \bold{W}}{dt}  \right)  \Delta (\bold{x}-\bold{x}^{\alpha}_i) +  \sum_i \left( \hat{\bold{v}}^1_i - \hat{\bold{v}}^2_i \right) \cdot \bold{f}_i  \Delta (\bold{x}-\bold{x}^{\alpha}_i) \\
& \hspace{2in} + \rho \bold{b} \cdot \bold{v} +  \rho \Gb \cdot \omegab \\ 
& \hspace{0.31in} =  \Lambda + \rho \bold{b} \cdot \bold{v} +  \rho \Gb \cdot \omegab \ ,
\end{split}
\end{equation}
where 
\begin{equation}\label{eq:internal_heat}
\Lambda=\sum_{i, \alpha} \left[ \frac{\bold{p}^{\alpha}_i}{m^{\alpha}_i} - \hat{\bold{v}}_i^\alpha \right] \cdot \left(  - \zeta \frac{\bold{p}^{\alpha} _i}{m_i} +  \sqrt{2\kt \zeta}\, \frac{\mathrm{d}\bold{W}}{\mathrm{d}t}  \right)  \Delta (\bold{x}-\bold{x}^{\alpha}_i)+\sum_{i, \alpha} \left( \hat{\bold{v}}^1_i - \hat{\bold{v}}^2_i \right) \cdot \bold{f}_i \Delta (\bold{x}-\bold{x}^{\alpha}_i) \ ,
\end{equation}
and $\rho \bb$ and $\rho \Gb$ are the body forces and the body torques given by \eqref{eq:body_force} and \eqref{eq:body_torque} respectively.

Using \eqref{eq:heat_5} and \eqref{eq:internal_heat}, the total energy balance at the macroscale \eqref{eq:fourth_energy_deriv} can be obtained as 
\begin{equation}
\label{eq:energy_final}
 \rho \dot{{e}}= -\frac{\partial}{\partial \bold{x}} \cdot \bold{J}_{\text{q}} + \frac{\partial}{\partial \bold{x}} \cdot \left( \bold{T}^T \bold{v} \right)+\frac{\partial}{\partial \bold{x}} \cdot \left( \bold{C}^T \boldsymbol{\omega} \right)-\frac{1}{2}\frac{\partial}{\partial \bold{x}} \cdot \left( \bold{Y}:  \boldsymbol{\omega} \otimes \boldsymbol{\omega}  \right)+ \Lambda  + \rho \bold{b} \cdot \bold{v}+\rho \bold{G}\cdot \boldsymbol{\omega}\ .
\end{equation}

Equation \eqref{eq:energy_final} can be considered as a generalization of the balance of energy from microscopic dynamics as originally conceived by Irving and Kirkwood \cite{irving1950statistical}, where only the effects of linear momentum were considered. It can be seen from \eqref{eq:energy_final} that the extension to include internal spin effects lead to additional terms corresponding to the rate of work or power from spin stresses $\Cb$ and body torques $\Gb$. Moreover, the effect of transport of moment of inertia due to the existence of moment of inertia flux is explicit in the balance of energy, in addition to being implicitly part of the definition of total energy $\rho e$. Importantly, the existence of spin affects the contributions for the heat flux in comparison to the original expression for heat flux derived by Irving and Kirkwood \cite{irving1950statistical}. One direct difference is the way in which the convection of energy by interaction forces and active forces occurs mainly by the momentum of the particles relative to the convective translational and rotational velocity of the continuum point $\xb$. However, it is interesting to see that the convection of energy through kinetic and potential energy is still affected by means of the momentum relative to the translational velocity of the continuum point. It is unclear to us at this moment physically why there exist two distinct modes of convection for interaction and energetic terms. 

Lastly, the balance of energy \eqref{eq:energy_final} includes a term $\Lambda$ that can be interpreted as the source of heat with contributions from both the friction from the bath and associated thermal forces, and the active torques. This shows how the bath and the active rotations can appear as internal sources of energy changes when viewed from a coarse-grained perspective. 

\subsubsection{Balance of Internal Energy}
In what follows, we use the decomposition of the total energy given in \eqref{eq:total_energy_short} and derive the balance of internal energy. We start by taking the time derivative of \eqref{eq:total_energy_short} which can be written as:
\begin{equation}
\label{eq:internal_energy1}
\rho \dot{{e}}=\rho \dot{\epsilon}+\rho \bold{v} \cdot \dot{\vb} +\frac{1}{2}\left( \dot{\bold{I}}+\bold{I} \frac{\partial}{\partial \bold{x}} \cdot \bold{v} \right) \boldsymbol{\omega} \cdot \boldsymbol{\omega}+\bold{I}\boldsymbol{\omega}\cdot \dot{\boldsymbol{\omega}} \ .
\end{equation}

Taking the dot product of the balance of linear momentum with the velocity vector $\vb$, equation \eqref{eq:mom_bal_final} yields 
\begin{equation}
 \rho \bold{v} \cdot \dot{\vb} = \frac{\partial}{\partial \bold{x}} \cdot \left( \bold{T}^T \bold{v} \right)  - \bold{T} : \nabla \bold{v} + \rho \bold{b} \cdot \bold{v} \ ,
\end{equation}
where use is made of the identity
\begin{equation}
\label{eq:tensor_identity}
\left( \frac{\partial}{\partial \bold{x}} \cdot \bold{A} \right) \cdot \bold{b} = \frac{\partial}{\partial \bold{x}} \cdot \left( \bold{A}^T \bold{b} \right) - \bold{A} : \nabla \bold{b} \ ,
\end{equation}
with $\Ab$ and $\bb$ being any arbitrary tensor and vector respectively.

Taking the total time derivative of the equation \eqref{eq:def_omega} corresponding to the definition of the rotational velocity of the continuum point yields 
\begin{equation}
\label{eq:deriv_rotational_vel_def}
\rho \dot{\bold{M}} - \rho \Big( \frac{\partial}{\partial \bold{x}} \cdot \bold{v} \Big) \bold{M} = \dot{\bold{I}}\boldsymbol{\omega} + \bold{I}\dot{\boldsymbol{\omega}} \ ,
\end{equation}
where use is made of \eqref{eq:mass_IK}. Taking the dot product of \eqref{eq:deriv_rotational_vel_def} with $\boldsymbol{\omega}$ and rearranging the terms yields
\begin{equation}\label{eq:int_1}
\rho \dot{\bold{M}} \cdot \boldsymbol{\omega} = \Big( \dot{\bold{I}} + \bold{I} \frac{\partial}{\partial \bold{x}} \cdot \bold{v} \Big) \boldsymbol{\omega} \cdot \boldsymbol{\omega} + \bold{I} \boldsymbol{\omega} \cdot \dot{\boldsymbol{\omega}} \ ,
\end{equation}
using \eqref{eq:def_omega}. Substituting \eqref{eq:final_spin} for the left hand side of \eqref{eq:int_1} yields
\begin{equation}
 \label{eq:intertia_terms}
 \left( \dot{\bold{I}} + \bold{I} \frac{\partial}{\partial \bold{x}} \cdot \bold{v} \right) \boldsymbol{\omega} \cdot \boldsymbol{\omega} + \bold{I} \boldsymbol{\omega} \cdot \dot{\boldsymbol{\omega}} = \frac{\partial}{\partial \bold{x}} \cdot \left( \bold{C}^T \boldsymbol{\omega} \right) - \bold{C}: \nabla \boldsymbol{\omega} + \rho \bold{G} \cdot \boldsymbol{\omega} + \bold{A} \boldsymbol{\omega} \ .
 \end{equation}
 Combining \eqref{eq:intertia_terms}, \eqref{eq:moi_bal_final}, and \eqref{eq:energy_final}, yields the balance of internal energy as
  \begin{equation}
 \begin{split}
 \label{eq:final_internal_energy_balance}
 \rho \dot{\epsilon} = -\frac{\partial}{\partial \bold{x}} \cdot \bold{J}_{\text{q}} + \bold{T} : \nabla \bold{v} + \bold{C} : \nabla \boldsymbol{\omega} - \bold{A} \cdot \boldsymbol{\omega} + \Lambda -  \frac{ \bold{Y} :\nabla (\boldsymbol{\omega} \otimes \boldsymbol{\omega} )}{2} \ ,
 \end{split}
 \end{equation}
 where we have used the following tensor calculus identity 
 \begin{equation}
 \frac{ \bold{Y} :\nabla (\boldsymbol{\omega} \otimes \boldsymbol{\omega} )}{2} \equiv \frac{1}{2} \frac{\partial}{\partial \bold{x}} \cdot \left( \bold{Y} : \boldsymbol{\omega} \otimes \boldsymbol{\omega} \right) + \frac{1}{2} \left( \frac{\partial}{\partial \bold{x}} \cdot \bold{Y} \right) \boldsymbol{\omega} \cdot \boldsymbol{\omega} \ . 
 \end{equation}

\section{Conclusion}

The coarse graining procedure presented in this paper can be considered as a generalization of the Irving and Kirkwood procedure to systems with internal rotational degrees of freedom. The expressions for the stress and couple stress tensor and the heat flux vector and their dependence on the active forces (or torques) may lead to a better understanding of the novel mechanical and rheological properties found in active matter systems.  Specifically, it is of interest to understand the effects of the active forces on the resulting effective transport coefficients such as viscosities and thermal conductivities. Having the expressions for the stress, couple stress, and heat flux vectors in terms of the molecular variables may facilitate calculations of the transport coefficients provided there exist Green-Kubo relations for these out of equilibrium systems. Finally, we note that our coarse-graining procedure and equations are generalizable to other active models such as self-propelled active Brownian particles.

{\center{\textbf{Acknowledgements}}}

The authors would like to thank Fr\'ed\'eric van Wijland, Steve Granick, Michael Hagan, David Limmer, Robert Jack, Michael R. DeWeese, and Panayiotis Papadopoulos for useful discussions.  
KKM acknowledges support from a National Institutes of Health Grant R01-GM110066. 
He is also supported by Director, Office of Science, Office of Basic Energy Sciences, Chemical Sciences Division, of the ~U.~S.~Department of Energy under contract No. DE-AC02-05CH11231. 
DM acknowledges support from the U. S. Army Research Laboratory and the U. S. Army Research Office under contract W911NF-13-1-0390. 
KK acknowledges support from an NSF Graduate Research Fellowship.


\begin{thebibliography}{10}

\bibitem{irving1950statistical}
J.~H. Irving and J.~G. Kirkwood.
\newblock The statistical mechanical theory of transport processes. iv. the
  equations of hydrodynamics.
\newblock {\em J. Chem. Phys.}, 18:817--829, 1950.

\bibitem{Ramaswamy_2010}
S.~Ramaswamy.
\newblock The mechanics and statistics of active matter.
\newblock {\em Ann. Rev. Condens. Matter Phys.}, 1:323--345, 2010.

\bibitem{Romanczuk_2012}
P.~Romanczuk, M.~Bar, W.~Ebeling, B.~Lindner, and L.~Schimansky-Geier.
\newblock Active brownian particles from individual to collective stochastic
  dynamics.
\newblock {\em Eur. Phys. J. Special Topics}, 202:1--162, 2012.

\bibitem{Marchetti_2013}
M.~C. Marchetti, J.~F. Joanny, S.~Ramaswamy, T.~B. Liverpool, J.~Prost, M.~Rao,
  and R.~Aditi Simha.
\newblock Hydrodynamics of soft active matter.
\newblock {\em Rev. Mod. Phys.}, 85:1143--1189, 2013.

\bibitem{Yeomans_2014}
J.~M. Yeomans, D.~O. Pushkin, and H.~Shum.
\newblock An introduction to the hydrodynamics of swimming microorganisms.
\newblock {\em Eur. Phys. J. Special Topics}, 223:1771, 2014.

\bibitem{Menzel_2015}
A.~M. Menzel.
\newblock Tuned, driven, and active soft matter.
\newblock {\em Phys. Rep.}, 554:1, 2015.

\bibitem{Bechinger_2016}
C.~Bechinger, R.~Di Leonardo, H.~L{\"o}wen, C.~Reichhardt, G.~Volpe, and
  G.~Volpe.
\newblock Active particles in complex and crowded environments.
\newblock {\em Rev. Mod. Phys.}, 88:045006, 2016.

\bibitem{Tailleur_2008}
J.~Tailleur and M.~E. Cates.
\newblock Statistical mechanics of interacting run-and-tumble bacteria.
\newblock {\em Phys. Rev. Lett.}, 100:218103, 2008.

\bibitem{Fily_2012}
Y.~Fily and M.~C. Marchetti.
\newblock Athermal phase separation of self-propelled particles with no
  alignment.
\newblock {\em Phys. Rev. Lett.}, 108:235702, 2012.

\bibitem{Buttinoni_2013}
I.~Buttinoni, J.~Bialk{\'e}, F.~K{\"u}mmel, H.~L{\"o}wen, C.~Bechinger, and
  T.~Speck.
\newblock Dynamical clustering and phase separation in suspensions of
  self-propelled colloidal particles.
\newblock {\em Phys. Rev. Lett.}, 110:238301, 2013.

\bibitem{Redner_2013}
G.~S. Redner, M.~F. Hagan, and A.~Baskaran.
\newblock Structure and dynamics of a phase-separating active colloidal fluid.
\newblock {\em Phys. Rev. Lett.}, 110:055701, 2013.

\bibitem{Mognetti_2013}
B.~M. Mognetti, A.~Saric, S.~Angioletti-Uberti, A.~Cacciuto, C.~Valeriani, and
  D.~Frenkel.
\newblock Living clusters and crystals from low-density suspensions of active
  colloids.
\newblock {\em Phys. Rev. Lett.}, 111:245702, 2013.

\bibitem{Cates_2015}
M.~E. Cates and J.~Tailleur.
\newblock Motility-induced phase separation.
\newblock {\em Ann. Rev. Condens. Matter Phys.}, 6:219--244, 2015.

\bibitem{Wu_2000}
X.-L. Wu and A.~Libchaber.
\newblock Particle diffusion in a quasi-two-dimensional bacterial bath.
\newblock {\em Phys. Rev. Lett.}, 84:3017--3020, 2000.

\bibitem{Leptos_2009}
K.~C. Leptos, J.~S. Guasto, J.~P. Gollub, A.~I. Pesci, and R.~E. Goldstein.
\newblock Dynamics of enhanced tracer diffusion in suspensions of swimming
  eukaryotic microorganisms.
\newblock {\em Phys. Rev. Lett.}, 103:198103, 2009.

\bibitem{banerjee2017odd}
D.~Banerjee, A.~Souslov, G.~A. Abanov, and V.~Vitelli.
\newblock Odd viscosity in chiral active fluids.
\newblock {\em arXiv:1702.02393}, 2017.

\bibitem{lopez2015}
H.~M. Lopez, J.~Gachelin, C.~Douarche, H.~Auradou, and E.~Clement.
\newblock Turning bacteria suspensions into superfluids.
\newblock {\em Phys. Rev. Lett.}, 115(028301), 2015.

\bibitem{Narayan_2007}
V.~Narayan, S.~Ramaswamy, and N.~Menon.
\newblock Long-lived giant number fluctuations in a swarming granular nematic.
\newblock {\em Science}, 317:105, 2007.

\bibitem{bialke2015negative}
J.~Bialk{\'e}, J.~T. Siebert, H.~L{\"o}wen, and T.~Speck.
\newblock Negative interfacial tension in phase-separated active brownian
  particles.
\newblock {\em Phys. Rev. Lett.}, 115:098301, 2015.

\bibitem{Kaiser_2014}
A.~Kaiser, A.~Peshkov, A.~Sokolov, B.~ten Hagen, H.~L{\"o}wen, and I.~S.
  Aronson.
\newblock Transport powered by bacterial turbulence.
\newblock {\em Phys. Rev. Lett.}, 112:158101, 2014.

\bibitem{Mallory_2014_PRE_II}
S.~A. Mallory, A.~Caric, C.~Valeriani, and A.~Cacciuto.
\newblock Anomalous thermomechanical properties of a self-propelled colloidal
  fluid.
\newblock {\em Phys. Rev. E}, 89:052303, 2014.

\bibitem{Takatori_2014_PRL}
S.~C. Takatori, W.~Yan, and J.~F. Brady.
\newblock Swim pressure and stress generation in active matter.
\newblock {\em Phys. Rev. Lett.}, 113:028103, 2014.

\bibitem{Solon_2015}
A.~P. Solon, Y.~Fily, A.~Baskaran, M.~E. Cates, Y.~Kafri, M.~Kardar, and
  J.~Tailleur.
\newblock Pressure is not a state function for generic active fluids.
\newblock {\em Nature Phys.}, 11:673--678, 2015.

\bibitem{Winkler_2015}
R.~G. Winkler, A.~Wysocki, and G.~Gompper.
\newblock Virial pressure in systems of spherical active brownian particles.
\newblock {\em Soft Matter}, 11:6680--6691, 2015.

\bibitem{Speck_2016_PRE}
T.~Speck and R.~L. Jack.
\newblock Ideal bulk pressure of active brownian particles.
\newblock {\em Phys. Rev. E}, 93:062605, 2016.

\bibitem{Joyeux_2016}
M.~Joyeux and E.~Bertin.
\newblock Pressure of a gas of underdamped active dumbbells.
\newblock {\em Phys. Rev. E}, 93:032605, 2016.

\bibitem{Nikola_2016}
N.~Nikola, A.~P. Solon, Y.~Kafri, M.~Kardar, J.~Tailleur, and R.~Voituriez.
\newblock Active particles with soft and curved walls: Equation of state,
  ratchets, and instabilities.
\newblock {\em Phys. Rev. Lett.}, 117:098001, 2016.

\bibitem{Joyeux_2017}
M.~Joyeux.
\newblock Recovery of mechanical pressure in a gas of underdamped active
  dumbbells with brownian noise.
\newblock {\em Phys. Rev. E.}, 95:052603, 2017.

\bibitem{Marconi_2017_arXiv}
U.~M.~B. Marconi, C.~Maggi, and M.~Paoluzzi.
\newblock Pressure in an exactly solvable model of active fluid.
\newblock {\em arXiv:1705.02481}, 2017.

\bibitem{Fily_2017}
Y.~Fily, Y.~Kafri, A.~P. Solon, J.~Tailleur, and A.~Turner.
\newblock Mechanical pressure and momentum conservation in dry active matter.
\newblock {\em arXiv:1704.06499}, 2017.

\bibitem{Sandford_2017_arXiv}
C.~Stanford, A.~Grosberg, and J.-F. Joanny.
\newblock Pressure and flow of exponentially self-correlated active particles.
\newblock {\em arXiv:1705.01631}, 2017.

\bibitem{clausius1870}
R.~Clausius.
\newblock On a mechanical theory applicable to heat.
\newblock {\em Phil. Mag.}, 40:122--127, 1870.

\bibitem{Steffenoni2017}
S.~Steffenoni, G.~Falasco, and K.~Kroy.
\newblock Microscopic derivation of the hydrodynamics of
  active-brownian-particle suspensions.
\newblock {\em Phys. Rev. E}, 95(052142), 2017.

\bibitem{Dahler_1961}
J.~S. Dahler and L.~E. Scriven.
\newblock Angular monetum of continua.
\newblock {\em Nature}, 192:36--37, 1961.

\bibitem{Dahler_1963}
J.~S. Dahler and L.~E. Scriven.
\newblock Theory of structured continua i. general consideration of angular
  momentum and polarization.
\newblock {\em Proc. R. Soc. A}, 275:504, 1963.

\bibitem{deGroot_1984}
S.~R. de~Groot and P.~Mazur.
\newblock {\em Nonequilibrium thermodynamics}.
\newblock Dover, New York, 1984.

\bibitem{Stokes_1966}
V.~K. Stokes.
\newblock Couple stresses in fluids.
\newblock {\em Phys. Fluids}, 9:1709, 1966.

\bibitem{Stokes_1984}
V.~K. Stokes.
\newblock {\em Theories of fluids with microstructure}.
\newblock Springer-Verlag, New York, 1984.

\bibitem{Spencer_2004}
A.~J.~M. Spencer.
\newblock {\em Continuum mechanics}.
\newblock Dover, Mineola, New York, 2004.

\bibitem{lau2009fluctuating}
A.~W.~C. Lau and T.~C. Lubensky.
\newblock Fluctuating hydrodynamics and microrheology of a dilute suspension of
  swimming bacteria.
\newblock {\em Phys. Rev. E}, 80:011917, 2009.

\bibitem{Stark_2005}
H.~Stark and T.~C. Lubensky.
\newblock Poisson bracket approach to the dynamics of nematic liquid crystals:
  The role of spin angular momentum.
\newblock {\em Phys. Rev. E}, 72(5):051714, 2005.

\bibitem{Tsai2005}
J.~C. Tsai, F.~Ye, J.~Rodriguez, J.~P. Gollub, and T.~C. Lubensky.
\newblock A chiral granular gas.
\newblock {\em Phys. Rev. Lett.}, 94(21):214301, 2005.

\bibitem{van2016spatiotemporal}
B.~C. van Zuiden, J.~Paulose, W.~T.~M. Irvine, D.~Bartolo, and V.~Vitelli.
\newblock Spatiotemporal order and emergent edge currents in active spinner
  materials.
\newblock {\em Proc. Natl. Acad. Sci}, 113:12919--12924, 2016.

\bibitem{Mandal2017short}
D.~Mandal, K.~Klymko, and K.~K. Mandadapu.
\newblock Generalized hydrodynamics of active polar suspensions.
\newblock {\em arXiv:1706.02284}, 2017.

\bibitem{zwanzig2001nonequilibrium}
R.~Zwanzig.
\newblock {\em Nonequilibrium statistical mechanics}.
\newblock Oxford University Press, 2001.

\bibitem{mandadapu2012homogenization}
K.~K. Mandadapu, A.~Sengupta, and P.~Papadopoulos.
\newblock A homogenization method for thermomechanical continua using extensive
  physical quantities.
\newblock {\em Proc. R. Soc. A}, 468:1696--1715, 2012.

\bibitem{Noll_1955}
R.~B. Lehoucq and A.~Von Lilienfeld-Toal.
\newblock Translation of walter noll\textsc{\char13}s derivation of the
  fundamental equations of continuum thermodynamics from statistical mechanics.
\newblock {\em J. Elasticity}, 100:5--24, 2010.

\bibitem{Noll2_1955}
W.~Noll.
\newblock {\em J. Rat. Mech. Analysis}, 5:627--646, 1955.

\bibitem{Kranthi2009a}
K.~K. Mandadapu, R.~E. Jones, and P.~Papadopoulos.
\newblock Generalization of the homogeneous non-equilibrium molecular dynamics
  method for calculating thermal conductivity to multi-body potentials.
\newblock {\em Phys. Rev. E}, 80:047702--1--4, 2009.

\bibitem{Kranthi_thesis}
K.~K. Mandadapu.
\newblock Homogeneous non-equilibrium molecular dynamics methods for
  calculating the heat transport coefficient of solids and mixtures.
\newblock 2011.

\bibitem{Stillinger1985}
F.~H. Stillinger and T.~A. Weber.
\newblock Computer-simulation of local order in condensed phases of silicon.
\newblock {\em Phys. Rev. B}, 31:5262--5271, 1985.

\bibitem{molinero2008water}
V.~Molinero and E.~B. Moore.
\newblock Water modeled as an intermediate element between carbon and silicon.
\newblock {\em J. Phys. Chem. B}, 113(13):4008--4016, 2008.

\bibitem{limmer2011putative}
D.~T. Limmer and D.~Chandler.
\newblock The putative liquid-liquid transition is a liquid-solid transition in
  atomistic models of water.
\newblock {\em The Journal of chemical physics}, 135(13):134503, 2011.

\bibitem{KranthiSilicon}
K.~K. Mandadapu, R.~E. Jones, and P.~Papadopoulos.
\newblock A homogeneous non-equilibrium molecular dynamics method for
  calculating thermal conductivity with the three-body potential.

\bibitem{hatwalne2004rheology}
Y.~Hatwalne, S.~Ramaswamy, M.~Rao, and R.~A. Simha.
\newblock Rheology of active-particle suspensions.
\newblock {\em Phys. Rev. Lett.}, 92:118101, 2004.

\bibitem{landau1986theory}
L.~D. Landau, A.~M. Kosevich, L.~P. Pitaevskii, and E.~M. Lifshitz.
\newblock Theory of elasticity.
\newblock 1986.

\end{thebibliography}

\end{document}